# The 29th Research Conference on Information, Communication, & Internet Policy
## TPRC 2001

### FROM 2G TO 3G – THE EVOLUTION OF INTERNATIONAL CELLULAR STANDARDS

**By**

**Audrey N. Selian**

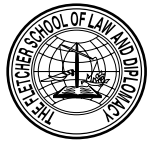

**Edward R. Murrow Center for International Communications Research**

**Fletcher School of Law and Diplomacy**

**Tufts University**

**September 2001**

**ACKNOWLEDGEMENTS**

I would like to thank Prof. Lee McKnight and members of the Murrow Center of Public Diplomacy at the Fletcher School of Law and Diplomacy of Tufts University, for their feedback and suggestions. This document was prepared in association with the International Telecommunication Union (ITU) in Geneva, Switzerland, and I would like to thank Dr. Tim Kelly, Ms. Lara Srivastava, Mr. Robert Shaw and Mr. Fabio Leite for their valuable guidance and advice in the preparation of this paper.

**TABLE OF CONTENTS:**

## TABLES AND FIGURES:

# 1 Introduction

Remarkable changes are taking place in the arena of mobile technologies, and the worldwide push toward $3^{rd}$ generation services is currently at the forefront of these transformations. Many questions surround the concept of 3G – not only in terms of what it means and what services it will offer, but also in terms of how to get there, which standard will be dominant, how long it will take to deploy, and whether it will be as lucrative as expected given the current rush of exorbitant spectrum fees.

This case study is designed to examine some of these questions about 3G from the analytical perspective of predecessor $2^{nd}$ generation technologies, and specifically that of GSM in Europe. The successful development and deployment of GSM over the past two decades is most significant, if one is to accept the hypothesis that 'experience counts' in the mobile arena. $3^{rd}$ generation mobile technologies must, after all, in some way be the result of an evolution from pre-existing 2G systems, whether this is because they are developed from overlays on $2^{nd}$ generation systems, or because operators deploying them must leverage pre-established 2G infrastructure or customer bases. The two are in many ways inextricably linked, and therefore examining one necessarily implies looking at the successes/shortcomings of the other.

Prior to the market liberalization of the 1990s, European telecom markets were firmly controlled by national governments and their respective PTT monopolists. Over the past decade, European telecommunications policy has been characterized by principles of market liberalization, harmonization of conditions of the regulatory framework, and the promotion of the European telecommunications industry. "GSM momentum" has been born of this environment, and is by far the biggest 2G system, with pan-European coverage and systems also installed in Asia, Australia, North America and more recently in South America.

The deployment of GSM is most aptly characterized by the commitment of twenty-six European national phone companies to standardize a system, and the working process responsible for this accomplishment has been deemed a great success worthy of replication. Essentially, those countries and firms involved realized the advantages of a cross-border standard and the amount of money and energy that can be wasted when competing for mobile technology 'world domination'.[1] Generally speaking, the story of the establishment of GSM is of interest to anybody studying the growth and trajectory of digital technology and its commercial applications. After all, as some have argued, the nature of digital economies implies that control over network evolution translates into control over the architecture of the digital marketplace."[2] The GSM case has proven that a hold over national networks has global economic ramifications.

Among the factors that helped to precipitate the creation of GSM, was the realization that localized solutions to the development of mobile communications would not be able to generate the economies of scale – from the R&D, production as well as distribution standpoints – necessary to attain very significant market penetration. With strides in the development of the realm of R&D came also the realization that only international market penetration goals could justify such extensive programs of investment. Long-term economic goals would be subjugated to the constraints of an unstandardized mobile communications sector, unless action could be taken to create some sort of consensus.

The existence of tremendous potential value in the network itself, following the logic of Metcalfe's Law and network economies, in addition to the value of scale economies in equipment markets, ensured that no government would lose out by agreeing to merely multilateral solutions when more widely cooperative institutional options were possible. After all, GSM was a network standard – not merely a product standard – and this had considerable significance in terms of the potential benefits to be derived from associated network externalities. Disharmony and the licensing of competing operators actually helped to make GSM a significant success in Europe: quality of service prior to GSM was low, and handsets were expensive. Thanks to a series of rather fortuitous market occurrences as well as to the efforts of Germany, the necessary impetus was provided to get GSM off the ground. European markets happened to open up to competition right around the time that the GSM standard was developed, resulting in a massive surge in demand for cellular phones. It is important to note that success came about in two parts: the initial interstate bargain, and the ensuing collaborative implementation once agreement was reached. The purpose of this paper is to examine the major factors surrounding and contributing to the creation (and success) of Europe's 2nd generation 'GSM' cellular system, and compare and contrast it to key events and recent developments in 3rd generation 'IMT-2000' systems.[3] The objective is to ascertain whether lessons from the development of one system can be applied to the other, and what implications 2G has for the deployment and assessment of 3G technologies.

---

[1] Andersson, Christoffer. <u>GPRS and 3G Wireless Applications</u> (New York, NY: Wiley Computer Publishing,, 2001), pp. 14-15.

[2] From a presentation given by Francois Bar called "*The Digital Economy in Comparative Perspective*", May 27, 1999 in Washington D.C. Referenced by Bach, David. "*International Cooperation and the Logic of Networks: Europe and the Global System for Mobile Communications (GSM)*". University of California E-conomy Project, Berkeley Roundtable on the International Economy (BRIE) – $12^{th}$ International Conference of Europeanists, Chicago, Illinois. March 30 – April 1, 2000, p.1.

## 1.1 The Generations of Mobile Networks

The idea of cell-based mobile radio systems appeared at Bell Laboratories in the United States in the early 1970s. However, mobile cellular systems were not introduced for commercial use until a decade later. During the early 1980's, analog cellular telephone systems experienced very rapid growth in Europe, particularly in Scandinavia and the United Kingdom. Today, cellular systems still represent one of the fastest growing telecommunications systems. During development, numerous problems arose as each country developed its own system, producing equipment limited to operate only within the boundaries of respective countries, thus limiting the markets in which services could be sold.

First-generation cellular networks, the primary focus of the communications industry in the early 1980's, were characterized by a few compatible systems that were designed to provide purely local cellular solutions. It became increasingly apparent that there would be an escalating demand for a technology that could facilitate flexible and reliable mobile communications. By the early 1990's, the lack of capacity of these existing networks emerged as a core challenge to keeping up with market demand. The first mobile wireless phones utilized analog transmission technologies, the dominant analog standard being known as "AMPS", (Advanced Mobile Phone System). Analog standards operated on bands of spectrum with a lower frequency and greater wavelength than subsequent standards, providing a significant signal range per cell along with a high propensity for interference.[4] Nonetheless, it is worth noting the continuing persistence of analog (AMPS) technologies in North America and Latin America through the 1980's.

Initial deployments of second-generation wireless networks occurred in Europe in the 1980's. These networks were based on digital, rather than analog technologies, and were circuit-switched. Circuit-switched cellular data is still the most widely used mobile wireless data service. Digital technology offered an appealing combination of performance and spectral efficiency (in terms of management of scarce frequency bands), as well as the development of features like speech security and data communications over high quality transmissions. It is also compatible with Integrated Services Digital Network (ISDN) technology, which was being developed for land-based telecommunication systems throughout the world, and which would be necessary for GSM to be successful. Moreover in the digital world, it would be possible to employ very large-scale integrated silicon technology to make handsets more affordable.

To a certain extent, the late 1980's and early 1990's were characterized by the perception that a complete migration to digital cellular would take many years, and that digital systems would suffer from a number of technical difficulties (i.e., handset technology). However, second-generation equipment has since proven to offer many advantages over analog systems, including efficient use of radio-magnetic spectrum, enhanced security, extended battery life, and data transmission capabilities. There are four main standards for 2G networks: Time Division Multiple Access (TDMA), Global System for Mobile Communications (GSM) and Code Division Multiple Access (CDMA); there is also Personal Digital Cellular (PDC), which is used exclusively in Japan.[5] (See Figure 1.1) In the meantime, a variety of 2.5G standards (to be discussed in Section 3.1) have been developed. 'Going digital' has led to the emergence of several major 2G mobile wireless systems.

---

[3] For the purposes of this paper, 'IMT-2000' is often considered in its European context, hence its interchangeability with the acronym 'UMTS' (Universal Mobile Telecommunications System).

[4] Guyton, James. *"Wireless Networks in Europe: A Three-Step Evolution"*. The Fletcher School of Law & Diplomacy, April 2000, p. 8.

[5] This system has put Japan in an awkward situation with an old system that was incompatible with all of the others; it has helped to jumpstart Japanese operators' aggressive pursuit of new technology and standards. In the late 1990's, cdmaOne began gaining ground in the Japanese market, increasing the pressure even more on the existing PDC operators. Andersson, Christoffer. GPRS and 3G Wireless Applications". (New York, N.Y.: Wiley Computer Publishing, 2001), p. 15.

**Figure 1.1: The 4 operational digital cellular technologies: Dec '00 (637 million users)**

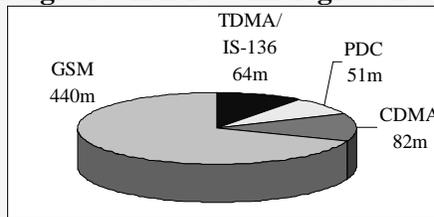

Source: *International Telecommunication Union*

TDMA was given an 'add-on' to create 'Digital AMPS' (D-AMPS), which facilitated the ability of handsets to switch between analog and digital operation. TDMA is the most widely used 2G technology in the western hemisphere (See Table 1.1) and is the base for GSM and PDC systems. Bits of voice are digitised and transmitted through an individual data channel, and then reconstructed at the other end of the channel to be converted back to voice.[6]

CDMAone (also referred to as IS-95), a solution that Qualcomm introduced in the mid 1990s, picked up toward the end of the decade. CDMA in general uses digital encoding and spread-spectrum techniques to let multiple users share the same channel; it differentiates users' signals by encoding them uniquely, transmitting through the frequency spectrum, and detecting and extracting the users' information at the receiving end. CDMA is noted to increase system capacity by about ten to fifteen times compared with AMPS, and by more than three times compared with TDMA. The industry recognizes CDMA as a superior air interface technology compared with that used in GSM/TDMA. However, what makes GSM popular is its successfully established international roaming feature.[7] Asia boasts a wide deployment of CDMA systems, thanks largely to Korea's investments in the technology; these systems, of course, represent the most advanced of second-generation technologies, providing much more reliable error recovery than TDMA counterpart alternatives.

GSM is a typical 2G system in that it handles voice efficiently, but provides limited support for data and Internet applications. Operators frequently point to GSM penetration levels of more than 50% in order to justify required investments in 3G licenses, network construction, and services development.[8] That the extent of the costs of deployment for 3G has rendered it a 'costly business' is a tremendous understatement. What sort of light could the GSM experience shed on the potential for acceptable ROI (returns on investment) for operators amidst this evolution? What key lessons have we learnt from GSM's time frame of deployment as well as its major drivers of success?

**Table 1.1: Regional Dominance of Current Wireless Technology Standards[9]**

| % of Total Subscribers 1999 | | | | |
|---|---|---|---|---|
| **Europe** | **North America** | **Latin American** | **Asia Pacific** | **Africa** |
| GSM: 89% | AMPS, other: 60% | AMPS, other: 55% | GSM: 35% | GSM: 88% |
| Other: 11% | TDMA: 27% | TDMA: 39% | CDMA: 14% | Other: 12% |
| | CDMA: 9% | CDMA: 5% | TDMA: 3% | |
| | GSM: 4% | GSM: 1% | Other: 48% | |

*Source:* ITU World Telecommunications Report 1999

---

[6] TDMA is a digital air interface that divides a single radio frequency channel into 6 unique time slots, allowing a number of users to access a single channel at one time without interference. By dividing the channel into slots, three signals (two time slots for each signal) can be transmitted over a single channel. In this way, TDMA technology (also referred to as ANSI-136), provides a 3 to 1 gain in capacity over analog technology. There are 115 million projected worldwide TDMA subscriber for 2001. For more information, see the *Universal Wireless Communications Consortium* website. (accessed July 2001) Link: http://www.uwcc.org/edge/tdma_faq.html.

[7] "*Generation Wireless*", Network Computing, Volume 12, Issue 12, June 11, 2001, p. 118.

[8] "*Wireless: Riding its luck into 3G*". Mobile Matters, February 2001, p.53.

[9] ITU World Telecommunications Development Report 1999.

The GSM standard clearly dominated the European market in 1999, with an 89% share.[10] (See Table 1.1) Today, Germany, the United Kingdom, Italy and France represent significant portions of subscribers relative to total European subscribers. (See Figure 1.2) GSM systems are based on technologies similar to TDMA, except for the fact that they operate at higher frequencies. Region by region, Europe, Asia Pacific and North America are experiencing a dramatic pace of expansion wherein GSM has become a dominant standard, with a high degree of extra services and ensuing popularity. However, with around 35 million customers already in 2000, China retains its position as the largest single GSM market in the world. Market penetration is reaching 70% in many developed GSM markets, with Finland and Italy expecting to be the first countries to reach 100%. In several Asia Pacific markets, the penetration of mobile wireless phones is overtaking that of fixed line phones.[11]

**Figure 1.2: World GSM Cellular Subscribers to June 2001**[12]

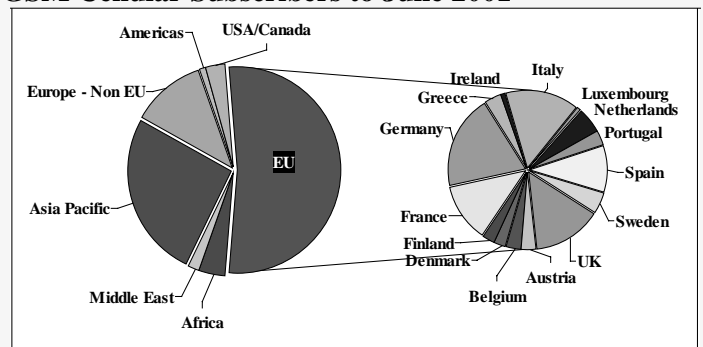

"GSM is now in more countries than McDonalds."[13] -Mike Short, Chairman of GSM MoU Association

Source: *http://www.gsmworld.com*

Carriers around Europe and Asia have gained a two-year lead in deploying 3G services; Japan is expected to see full 3G some time in 2001, while European deployment is anticipated sometime in 2002. According to the Strategis Group, 3G will not roll out in the United States until 2004."[14] Europe is expected to begin offering 3G products in 2002, followed by the U.S., which [optimistic] analysts predict, will likely launch 3G service between 2003 and 2005.[15] The third generation of mobile communications, as distinct from its predecessors, is likely to change many areas of social and economic activity, and is expected to unleash a wave of investment in the creation of new data-intensive services – the likes of which we can not yet aptly envisage in detail. It is not an exaggeration to expect many of these changes to be revolutionary, in a way that will likely be difficult, expensive and destructive, fundamentally affecting existing trends in the development of current technologies and the companies that support them (see L.McKnight and P.Vaaler's notion of "Creative Destruction")[16]. But they will also likely be liberating, rewarding and creative.

---

[10] Guyton, p. 9.

[11] "*GSM Approaches half a billion customers*", The GSM Association, Press Release - April 2000, p.1. (accessed October 2001) Link: http://www.gsmworld.com/news/press_releases_54.html.

[12] "*World Cellular Subscribers by Region: Subscriber Growth to June 01*". The GSM Association (accessed October 2001) Link: http://www.gsmworld.com/membership/graph4.html.

[13] Quote from Mr. Mike Short of The GSM MoU Association 1995-6. "*A Gaze into the Future*", The GSM Association (accessed October 2001) See Link: http://www.gsmworld.com /about/history_page17.html.

[14] Goodman, Peter S. "*A Push for More Frequencies: Wireless Firms Say They Can't Advance Until Government Opens Up the Airwaves*". Washington Post, February 28, 2001. P. G12.

[15] Hamblen, Mark. "*3G Wireless*", Feb. 21, 2000 (estimating that Japan will be the first to launch an advanced 3G system early in 2001, followed by Europe in 2002 and the U.S. between 2004 and 2005); see also "*Implementation of Section 6002(b) of the Omnibus Budget Reconciliation Act of 1993*", Annual Report and Analysis of Competitive Market Conditions With Respect to Commercial Mobile Services, FCC 00-289, at 38 (rel. Aug. 18, 2000).

## 2   A Look Back at GSM

### 2.1   GSM Technology

One of the most important conclusions from the early tests of the new GSM technology was that the new standard should employ Time Division Multiple Access (TDMA) technology. This ensured the support of major corporate players like Nokia, Ericsson and Siemens, and the flexibility of having access to a broad range of suppliers and the potential to get product faster into the marketplace. After a series of tests, the GSM digital standard was proven to work in 1988.

With global coverage goals in mind, being compatible with GSM from day one is a prerequisite for any new system that would add functionality to GSM. As with other 2G systems, GSM handles voice efficiently, but the support for data and Internet applications is limited. A data connection is established in just the same way as for a regular voice call; the user dials in and a circuit-switched connection continues during the entire session. If the user disconnects and wants to re-connect, the dial-in sequence has to be repeated. This issue, coupled with the limitation that users are billed for the time that they are connected, creates a need for packet data for GSM.

The digital nature of GSM allows the transmission of data (both synchronous and asynchronous) to or from ISDN terminals, although the most basic service support by GSM is telephony.[17] Speech, which is inherently analog, has to be digitized. The method employed by ISDN, and by current telephone systems for multiplexing voice lines over high-speed trunks and optical fiber lines, is Pulse Coded Modulation (PCM). From the start, planners of GSM wanted to ensure ISDN compatibility in services offered, although the attainment of the standard ISDN bit rate of 64 Kbit/s was difficult to achieve, thereby belying some of the limitations of a radio link. The 64 Kbit/s signal, although simple to implement, contains significant redundancy.

Since its inception, GSM was destined to employ digital rather than analog technology and operate in the 900 MHz frequency band. Most GSM systems operate in the 900 MHz and 1.8 GHz frequency bands, except in North America where they operate in the 1.9 GHz band. GSM divides up the radio spectrum bandwidth by using a combination of Time- and Frequency Division Multiple Access (TDMA/FDMA) schemes on its 25 MHz wide frequency spectrum, dividing it into 124 carrier frequencies (spaced 200 Khz apart). Each frequency is then divided into eight time slots using TDMA, and one or more carrier frequencies are assigned to each base station. The fundamental unit of time in this TDMA scheme is called a 'burst period' and it lasts 15/26 ms (or approx. 0.577 ms). Therefore the eight 'time slots' are actually 'burst periods', which are grouped into a TDMA frame, which subsequently form the basic unit for the definition of logical channels. One physical channel is one burst period per TDMA frame.[18]

---

[16]While the Internet economy has strong implications for business strategy, traditional economic dynamics still apply to firms doing business in the Internet (and, by extension, arguably in the wireless) environment. Contrary to the predictions of new-economy optimists, the Internet industry has not brought an end to the business cycle or created boundless opportunity for an unlimited number of new entrants. Companies will still have to compete, and those that emerge as successful must constantly respond to the changing conditions of their business environment. This is where Joseph Schumpeter's theory of creative destruction meets the 'Internet economy'. In the Internet world, old businesses and industries will be destroyed even more rapidly, and firms must learn to identify, cope with, encourage, and exploit this dynamic. From a summary of "Creative Destruction: Business Survival Strategies in the Global Internet Economy" (March 2001) published by L. McKnight and P. Vaaler, The Fletcher School of Law and Diplomacy. (accessed October 2001) See Link: http://www.ceip.org/files/events/mcknight.asp?pr=1&EventID=320.

[17] "The GSM group studied several speech coding algorithms on the basis of subjective speech quality and complexity (which is related to cost, processing delay, and power consumption once implemented) before arriving at the choice of a Regular Pulse Excited -- Linear Predictive Coder (RPE--LPC) with a Long Term Predictor loop. Basically, this is a method whereby information from previous samples, which tends not to change quickly, is applied to predict the current sample. Data can use either the transparent service, which has a fixed delay but no guarantee of data integrity, or a non-transparent service, which guarantees data integrity through an Automatic Repeat Request (ARQ) mechanism, but with a variable delay. The data rates supported by GSM are 300 bit/s, 600 bit/s, 1200 bit/s, 2400 bit/s, and 9600 bit/s." From Scourias, John. "*A Brief History of GSM*". University of Waterloo, 1994. (accessed October 2001) Link: http://kbs.cs.tu-berlin.de/~jutta/gsm/js-intro.html.

[18] Scourias, John. "*Overview of the Global System for Mobile Communications*". University of Waterloo, 1997. (accessed July 2001) Link: http://www.shoshin.uwaterloo.ca/ ~jscouria /GSM/gsmreport.html.

The development of standards and systems spans well beyond the technical realm and often into the political; this is best exemplified by what happened with GSM. Shortly after the suitability of TDMA for GSM was determined, a political battle erupted over the question of whether to adopt a wide-band or narrow-band TDMA solution. Whereas France and Germany supported a wide-band solution, the Scandinavian countries favored a narrow-band alternative. These governmental preferences were clearly a reflection of the respective countries' domestic equipment manufacturers as German and French manufacturers SEL and Alcatel had invested substantially into wide-band technology, whereas their Scandinavian counterparts Ericsson and Nokia poured resources into the narrow-band alternative. Italy and the UK, in turn, were the subjects of intense lobbying on behalf of the two camps with the result of frequently changing coalitions.[19] The culmination of this controversy between the two camps was a CEPT (Conference des Administrations Europeans des Posts et Telecommunications) Meeting in Madeira in February 1987. The Scandinavian countries finally convinced Italy, the UK and a few smaller states of the technical superiority of narrow-band technology and left Germany and France as the only proponents of the wide-band alternative. Since CEPT followed purely intergovernmental procedures, however, decisions had to be taken unanimously, and Germany and France were able to veto a decision that would have led to the adoption of narrow-band TDMA as the technology underlying the GSM project.

A unique feature of GSM is the Short Message Service (SMS), which has achieved wide popularity as what some have called the unexpected 'killer application' of GSM. SMS is a bi-directional service for sending short alphanumeric message in a store-and-forward process. SMS can be used both 'point-to-point' as well as in cell-broadcast mode. (Further information in Section 3.5) Supplementary services are provided on top of teleservices or bearer services, and include features such as, among others, call forwarding, call waiting, caller identification, three-way conversations, and call-barring.

The most novel and far-reaching feature of GSM is that it provides most of Europe's cellular phone users with a choice – choice of network and choice of operator. Also, international roaming was and continues to be the cornerstone of GSM. For this to be possible, all networks and handsets have to be identical. With many manufacturers creating many different products in many different countries, each type of terminal has been put through a rigorous approval regime. However, at the time, no approval process was available, and it took nearly a year before the handheld terminals were tested and fit for market entry.

Another of GSM's most attractive features is the extent to which its network is considered to be secure. All communications, both speech and data, are encrypted to prevent eavesdropping, and GSM subscribers are identified by their Subscriber Identity Module (SIM) card (which holds their identity number and authentication key and algorithm). While the choice of algorithm is the responsibility of individual GSM operators, they all work closely together through the Memorandum of Understanding (MoU) (to be described in greater detail in section 2.2.2) to ensure security of authentication. This smartcard technology minimizes the necessity for owning terminals - as travellers can simply rent GSM phones at the airport and insert their SIM card. Since it's the card rather than the terminal that enables network access, feature access and billing, the user is immediately on-line.

## 2.2   The History of GSM

The Western European mobile wireless market has not been forged by market forces alone. Indeed as mentioned previously, the harmonization of standards and interoperability were due in large part to governmental efforts. These public sector influences carry over to the next generation of mobile cellular networks, as well as through the ITU's IMT-2000 initiative – which is embodied in UMTS movement in Europe.[20]

The GSM story began in the early 1980's, when European countries struggled with no fewer than nine competing analog standards, including Nordic Mobile Telephony (NMT), Total Access Communications Systems (TACS), and so on. In order to put the rise of GSM in context, it is important to note that the climate of economic liberalization and opening up of new markets in Asia, Latin American and Eastern Europe helped boost analog system subscriber numbers throughout the 1990's. The roll-out of a multi-national global communications standard faced several formidable barriers. Operators were concentrating on new methods for expanding old analog networks, using methods like NAMPS (Narrowband Advanced Mobile Phone Service) by Motorola; unsurprisingly, there was resistance to the prospects of a digital launch.

---

[19] Bach, David. "*International Cooperation and the Logic of Networks: Europe and the Global System for Mobile Communications (GSM)*". University of California E-conomy Project, Berkeley Roundtable on the International Economy (BRIE) – 12[th] International Conference of Europeanists, Chicago, Illinois. March 30 – April 1, 2000, p.5.

[20] Guyton, p. 77.

Pan-European roaming was nothing more than a distant dream at that point, and capacity was a particularly difficult issue. Europeans recognized the need for a completely new system – a system that could accommodate an ever-increasing subscriber base, advanced features and standardized solutions across the continent. Because of the shortcomings and incompatibility issues associated with analog systems, a completely new digital solution was instituted. The new standard, Groupe Spéciale Mobile (GSM), was built as a wireless counterpart of the land-line Integrated Services Digital Network (ISDN) system. Although GSM initially stood for 'Groupe Spéciale Mobile', named after the study group that created it, the acronym was later changed to refer to 'Global System for Mobile communications'. This transition as well as other key aspects of GSM history are elaborated upon in subsequent sections.

**Table 2.1: Timeline of the development of GSM[21]**

| *Year* | *Events* |
|--------|----------|
| *1982* | CEPT establishes a GSM group in order to develop the standards for a pan-European cellular mobile system |
| *1985* | Adoption of a list of recommendations to be generated by the group |
| *1986* | Field tests were performed in order to test the different radio techniques proposed for the air interface |
| *1987* | TDMA is chosen as access method (in fact, it will be used with FDMA) Initial Memorandum of Understanding signed by the telecommunication operators (representing 12 countries) |
| *1988* | Validation of the GSM system |
| *1989* | The responsibility of the GSM specifications is passed to the ETSI |
| *1990* | Appearance of the phase I of the GSM specifications |
| *1991* | Set date for the 'official' commercial launch of the GSM service in Europe |
| *1992* | Actual launch of commercial service, and enlargement of the countries that signed the GSM – MoU > Coverage of Larger cities / airports |
| *1993* | Coverage of main roads GSM services start outside Europe |
| *1995* | Phase II of the GSM specifications Coverage of rural areas |

*Source*: The ITU and "An Overview of the GSM System".

### 2.2.1    Conference Des Administrations Europeans des Posts et Telecommunications (CEPT)

As soon as it became apparent that long-term economic goals in Europe had to be addressed, the CEPT was formed in 1982 by the "Conference Des Administrations Europeans Des Posts et Telecommunications" to address sector needs. The majority of CEPT's membership was comprised of state monopolies that were accustomed to considering their own national interests as primary objective. Nonetheless, at that time, awareness of the fact that the new industry's economic future relied on high levels of pan-European co-operation was tremendously important. Before CEPT formally launched the GSM project in 1982, cooperation on analog standards for mobile communications in Europe had been attempted between France and the UK, and France and Germany respectively. However, simultaneous efforts by national governments to protect their own industries frequently interfered with the realization of gains from cooperation. In the end, neither of the two projects was successful, and unilateral solutions in each of the larger European states left the European market fragmented, and networks incompatible with one another.

---

[21]Gozalvez Sempere, Jose. *"An Overview of the GSM System"*. Communications Division, Department of Electronic and Electrical Engineering, University of Strathclyde, Glasgow, Scotland. May 22, 2001 (accessed August 2001). Link: http://www.comms.eee.strath.ac.uk/ ~gozalvez/gsm/gsm.html, p. 3.

European PTT representatives were put in a position wherein exploration of the feasibility of multilateral cooperation was unavoidable. As it was, the existing analog mobile systems in place were totally incompatible with one another and limited to the extent of the respective national jurisdictions. The CEPT subsequently established the 'Groupe Spéciale Mobile' (GSM), to develop the specification for a pan-European mobile communications network capable of supporting the scores of subscribers who were projected to be likely customers of mobile communications services in the future. The standardized system was to meet a few criteria: spectrum efficiency, international roaming, low mobile and base stations costs, good subjective voice quality, compatibility with other systems such as ISDN, and the ability to support new services.[22]

### 2.2.2    The European Commission and the Memorandum of Understanding

"… the political process that enabled GSM featured pivotal supranational leadership in the form of European Commission initiatives in a domain that has traditionally been dominated by national players."[23]    A close examination of the emergence of GSM and its characteristics reveals that the critical 10-15 year period during which it emerged was characterized by a systematic process of thought leadership that served to challenge what would otherwise have been a well-engrained 'status quo' in the telecommunications sector. Certainly, the importance of the political undercurrents surrounding these events cannot be overstated; the implications of the 1984 endorsement by the European Commission (EC) of the GSM project are still in evidence today. In 1985, a small group of countries including France, West Germany, and Italy, together determined in an agreement for the development of GSM, that digital technology would become the future of global mobile wireless communication; the United Kingdom joined in the following year.

By the mid-1980's, pressure from countries like France and West Germany encouraged the Commission of the European Communities to outline the situation to the Heads of Member States at a meeting in December 1986. The GSM Permanent Nucleus (headquartered in Paris) was thereby formed to assume overall responsibility for coordinating the development of GSM, and Stephen Temple of the UK's Department of Trade and Industry was charged with the task of drafting the first Memorandum of Understanding (MoU). On September 7, 1987, network operators from thirteen countries signed an MoU in Copenhagen. There were 15 signatories in total: France, Germany, Italy, Sweden, Norway, Denmark, Finland, Spain, the Netherlands, Belgium, Portugal, Ireland, the DTI and two independent operators (Cellnet and Racal-Vodafone) from the UK.[24]  It was designed to forge the commercial agreement necessary between potential operators, so that commitments could officially made to implement the standard by a particular date. Without it, no network would have been established, no terminals would have been developed, and no service could have come into existence.

The "MoU" has come to represent GSM's main governing body and currently consists of 210 contributing members from 105 countries. The MoU's basic task is to establish internationally-compatible GSM networks in member countries, and to provide a mechanism to allow for cooperation between operators with respect to commercial, operational and technical issues. Generally, the GSM "MoU Plenary" meets every four months, and allows member organisations to discuss the direction in which GSM should develop, and to examine revisions and improvements to standard GSM MoU documents. The MoU includes members that operate GSM networks at 900 MHz (GSM 900), and also at the higher 1,800 MHz frequency (DCS 1800) and now 1,900 MHz (PCS). There are also a number of special interest groups representing operators groups by geographical location or technology. At each Plenary session, the chairpersons of various working groups bring members up to date with latest developments. These working groups examine issues such as international roaming, harmonization of tariff principles, global marketing, accounting and billing procedures, legal and regulatory matters, time scales for the procurements and deployment of systems, etc. Proposals are voted upon, with the number of votes allocated to a member dependent on factors like 'number of subscribers' or 'GDP'.[25]

In 1987, the Commission issued a "Green Paper" on the development of the common market for telecommunications services and equipment, emphasizing the crucial importance of a 'technically advanced,

---

[22] Gozalvez Sempere, Jose, p. 3.

[23] Bach, p.1.

[24]"*The Memorandum of Understanding*". GSM Association website. (accessed October 9, 2001) Link: http://www.gsmworld.com/about/history_page7.html.

[25] "*The GSM MoU: How It Works*". (accessed July 2001) Link: http://www.cellular.co.za/gsm-mou.htm.

Europe-wide, low-cost telecommunications network' for the competitiveness of the European economy. The Green Paper outlined the Commission's challenge to PTT dominance of European telecom markets by suggesting 'Community-wide' competition in the areas of network equipment, terminals, and communication services.[26] The result was a recommendation and a Directive, which between them laid and reinforced the political foundations for the development of GSM, and which called for a launch of a limited set of services by 1991. The Directive ensured that every Member State would reserve the 900 MHz frequency blocks required for the rollout program. Although these were somewhat smaller than the amount advocated by the CEPT, the industry had finally achieved the political support it needed to advance its objectives. By 1987 the critical "GSM Directive"[27] was created, the purpose of which was to ensure the goal of frequency harmonization across Europe's member states such that the goal of pan-European roaming could be achieved.

In sum, Europe's success mainly reflects a decision made a decade ago to back the pan-European GSM as the digital standard for mobile telephony; the European Commission was successfully able to leverage its institutional authority by stressing the link between the creation of a pan-European digital standard to issues of large European market integration.[28] This, coupled with EU-backed regulatory changes that mandated the licensing of competing GSM networks in all EU member states, led to the digital mobile telephony boom in Europe.

### 2.2.3 European Telecommunications Standards Institute (ETSI)

In 1989, the European Telecommunications Standards Institute (ETSI) was created in order to take responsibility for specification development from the GSM Permanent Nucleus. ETSI had a unique organizational structure that accorded equal status to administrators, operators and manufacturers; this equilibrated terrain had a considerable impact on the speed of development. Whereas CEPT was primarily a brokerage table for national governments and their PTT representatives, ETSI was an institutional actor in its own right, capable of concentrating the support of all relevant parties behind a project like GSM. It was this combination of a co-operative environment and improved resources that enabled the majority of Phase I of the GSM 900 specifications to be published in 1990.

This said, however, it was also important to note the considerable influence of EU institutions on ETSI's operations as well as on the implementation of standards, even though ETSI itself (like the CEPT) is formally a body independent of the European Union. The institutional arrangement gives EU institutions three ways of affecting ETSI's standardization efforts as well as standards implementation. The European Commission can provide ETSI with voluntary contributions to support the development of particular standards that it deems necessary for market competitiveness, and can also prevent the adoption of standards that may be desired by some members if it believes that those standards might inhibit the flow of trade. Most importantly, a Council of Decisions of December 22, 1986 "on standardization in the field of information and telecommunications, requires EU members and their telecommunications administrations to use official European standards in public procurements."[29]

### 2.2.4 The "Frequency Band" Obstacle Course

A series of developments regarding the frequency bands upon which such technology could work created an interesting 'obstacle course' through which GSM was to develop. In 1989, the UK Department of Trade and Industry published a discussion document called "*Phones on the Move*", which advocated the introduction of mass-market mobile communications using new technology and operating in the 1800 MHz-frequency band. The UK government licensed two operators to run what became known as Personal Communications Networks (PCN), which operated at the higher frequency, giving PCN operators virtually unlimited capacity. Previously designated bands at 900 MHz were far more limited, and the GSM community began to feel somewhat under threat. Ironically enough, the UK's PCN turned out to be more of an opportunity than a threat in the end. The new operators decided to utilize the GSM specification - slightly modified because of the higher frequency - and the development of what became known as 'DCS 1800' was carried out by the ETSI in parallel with GSM standardization. In fact, in 1997 'DCS 1800' was renamed 'GSM 1800' to reflect the affinity between the two technologies.

---

[26] Bach, p.6.

[27] "*Council Directive of June 25, 1987 on the frequency bands to be reserved for the coordinated introduction of public pan-European cellular digital land-based mobile communications in the Community*." Official Journal of the European Communities. No L 196/85. (87/372/EC) June 1987. (accessed October 1, 2001) Link: http://145.18.106.100/doc/telecomrecht/eu/en/87_372_EEC.pdf.

[28] Bach, p. 9.

[29] Ibid, p.12.

### 2.2.5   The Conclusion of the Interstate Bargain

The shift of the responsibility for GSM from the bargaining table of CEPT towards ETSI epitomizes the conclusion of the interstate bargain and the deliberate move toward the task of implementation. From this point on, governments or the PTT representatives and the national champions that they backed, were no longer the primary actors in the standardization process. Rather, a multitude of actors, akin to the diverse membership of ETSI, and the European Commission, moved into the spotlight.

The case of GSM, apart from the general complexity of the issue, was further complicated by the fact that the actors involved in the process changed considerably over time. While international deliberations began on the level of the PTT representatives, the final bargain was struck by national governments. Supranational institutions and private corporations had played key roles even before the general agreement was reached, but their importance grew substantially once it came to implementing the framework, determining technical specifications and rolling-out service.[30] Adaptation on the national level led states to explore new means to achieve their goals of promoting domestic industry, while simultaneously securing benefits for consumers.

### 2.2.6   The Launch

The launch of GSM took place in the latter part of 1992, with the first GSM digital cellular network going 'live' in Finland in 1991; Finland and Germany (in 1992) were among the first European countries to launch. Germany, specifically, was known as "a main driver of European GSM cellular penetration"[31] through the early 1990's. In early 1992, only three or four GSM networks had launched. Within seven years, GSM networks had over 50 million subscribers in Europe. By comparison, it took fixed networks nearly 50 years to acquire the same number of subscribers worldwide, and about 15 years for the Internet to attract 50 million users worldwide.[32] Among the early runners were Finland (two operators), Germany (two operators), Denmark (two operators), Portugal (two operators), Sweden (three operators), Italy and France. On June 17, 1992 the first roaming agreement was signed between Telecom Finland and Vodafone in the UK, amidst great concern amongst operators mainly as a result of non-existent or interim type-approved handsets.

By 1993 the MoU boasted 70 members from 48 countries and 25 roaming agreements. Approximately one million people were now using the GSM network, with the next million already on the horizon. And, perhaps most significant of all, the Australian company Telstra had added its name to the growing MoU membership. After two years, GSM had expanded beyond Europe and Australia, establishing a presence in India, Africa, Asia and the Arab world. By June 1995, the MoU was formally registered as an Association in Switzerland[33], with 156 members serving 12 million customers in 86 countries.[34]

GSM (and its twin system operating at 1800 MHz, called DCS1800) was at this time perceived to be one of a number of new or revamped mobile services entering the market, though its 'presence in a crowd' of competing 2G systems would not undermine the critical role it was to play. This was true not only due to technological features, but to how it was introduced, which was contributing to the reorganization not only of the cellular market, but of the configuration of the telecommunications services industry across Europe. Operators in 1993 were piggy-backing 'local' digital services, with lower access and call charges, on the 900 MHz GSM networks.

---

[30] Bach, p.1.

[31] "*GSM Standard's Influence Spreads Worldwide*" <u>Mobile Phone News</u>, Phillips Business Information, Inc.  March 1, 1993.

[32] Bout, Dirk M., Daum, Adam, Deighton, Nigel, Delcroix, Jean-Claude, Dulaney, Ken, Green-Armytage, Jonathan, Hooley, Margot, Jones, Nick, Leet, Phoebe, Owen, Gareth, Richardson, Peter, Tade, David.  "<u>The Next Generation of Mobile Networks Poses a $100 Billion Challenge for Europe</u>", Note Number: R-11-5053, Gartner Group.  September 19, 2000.

[33] Status as an 'association' in Switzerland is designed for organizations that pursue non-profit objectives and engage in beneficial, scientific, cultural, political or social activities. However, many of the more important associations are formed to pursue economic goals, for instance, professional organizations and trade unions. Non-profit associations may, for the better attainment of their goals, carry on an industrial or commercial activity. Associations acquire the status of a separate legal entity as soon as the articles of association are drawn up.  From "*Types of Business Entities", Commercial Law Page*." Link: (accessed July 2001) http://geneva.ch/ genevaguidetypesbusinessentities.htm.

[34] "*The Memorandum of Understanding*". <u>The GSM Association</u>. (accessed October 2001) Link: http://www.gsmworld.com/about/history_page14.html.

Most of Europe's public telecommunication operators (PTOs) at this point in time were anxious for privatization and greater operational flexibility, hence the frequent separation of mobile divisions, allowing for bids for overseas licenses and work with private sector partners. In the years ahead, licensing administrations throughout the world would employ this system – which utilized modified GSM specifications – as a means of introducing further competition into the mobile market. The impact of such intensive competition was to shift mobile communications away from the business community and into the mass market. The number of cellular subscribers in western Europe, which had grown by roughly a third in each of the two previous years, increased by almost 50% in 1993.

### 2.2.7    The United States and the FCC

On the other hand, the US and Japan were generally perceived at this time as being somewhat 'left on the sidelines' in the drive to standardize mobile telephones. In 1994, the US Federal Communications Commission – in attempts to forge ahead with their own mobile cellular markets, auctioned large blocks of spectrum in the 1900 MHz band in the United States. The aim was to introduce digital mobile cellular networks to the country in the form of a new kind of mass market Personal Communications Service. Slowly, the market started gathering momentum as handsets became more widely available; in order to foster continued competitiveness, the FCC deliberately ensured that the personal communications services (PCS) licenses were neutral with regard to the type of technology to be employed.

1994 also saw the creation of the Mobile Green Paper, which presented a common approach in the field of mobile and personal communications in the European Union.

## 2.3    The GSM Market

### 2.3.1    The GSM Success Story

GSM was already beginning to be seen as a sort of distinctive success story by the mid-1990's. While there were at this time still more users in North America than in Europe, more than half the growth in Europe was to be derived from digital systems as opposed to the growth in the US, which still operated on Analog AMPS networks that were close to reaching full capacity. GSM by this time was perceived to have achieved economies of scale in making PCS handsets cheap enough to compete with analog networks; it was also considered by this point to be a proven "technology". 1995 saw the completion of the GSM Phase II standardization and a demonstration of fax, video and data communication via GSM. It also produced an adaptation of PCS 1900 to meet the opportunities created by the recent FCC auction in the USA. US cellular operators were expected to face competition in 1995-6 from PCS companies, using high frequencies at a similar part of the spectrum to those allocated to the UK cellular operators Mercury One-2-One and Orange, and E-Plus in Germany. 1996 was characterized also by liberalization of the mobile and satellite sectors.

New PCS operators in the U.S. also recognized the advantages of an open standard in creating a global, multi-vendor market for products. This had the advantage of making network deployment more cost-effective. Once the FCC had opened the door, the major GSM vendors rapidly developed a GSM variation customized for the 1900 MHz-frequency band. The US therefore appeared to be very interested in the DCS-1800 (GSM-compatible standard). In November 1995, American Personal Communications launched the first commercial GSM service in the US. At around the same time, Qualcomm (in precursor CDMA phases) was developing "spread spectrum technology", but no handsets had yet been developed. As an alternative to TDMA systems, it was perceived as somewhat of a 'risky proposition'. The threat of being left behind in the rapidly advancing GSM marketplace was too great. By May 1997, there were already 15 PCS 1900 (now GSM 1900) networks and over 400,000 users.[35]

In Europe, by 1997, one new customer was signing up to GSM networks every second, according to estimates from the GSM MoU Association, the global industry body that represents 239 international GSM network operators, regulators and administrators of 109 countries/areas. Customer totals for GSM had reached 44 million and were equivalent to 28% of the world mobile wireless market. In 1998, the EU Green Paper on Convergence was written, the purpose of which was to launch a debate on the regulatory implications of the convergence of the telecommunications, media and IT sectors, and to discuss options for future regulatory policy.

---

[35] "*Going Global*". The GSM Association (accessed October 2001) Link: http://www.gsmworld.com/about/history_page14.html.

By 1999, the positive effects of GSM's success were readily discernible on Europe's equipment vendors, network operators, system integrators, and software developers. Europe's vendors benefited from the economies of scale and efficiencies associated with the development of a stable technology platform. Companies like Nokia and Ericsson have been able to leverage their expertise in building GSM networks in Europe to sell their GSM infrastructure projects into emerging markets, such as those of eastern Europe where many telecomms operators have 'leapfrogged' wireline systems in favor of mobile wireless networks. These are often easier, quicker and cheaper to roll out. GSM network operators by 1999 delivered service to more than 200 million users, making it the most successful mobile wireless technology in the world; it had more than 400 million subscribers by the end of 2000, and has been adding about 10 million more each month.[36]

Looking ahead, it is possible that GSM systems are heading for quieter times unless something can stimulate more growth. In many countries, networks will soon be reaching the limits of what can be achieved without extensive, and environmentally intrusive investments in new wireless masts. Also, as saturation among users approaches, growth will slow. Therefore, network operators are bound to face some critically important strategic decisions. In the next two years, they must continue to focus on satisfying rapid customer demand for mobile voice services and on meeting the basic customer needs of coverage, capacity and customer service. Those in countries with high mobile phone penetration rates will lead the way in developing services featuring data as well as voice. It is likely that acquisitions and alliance activities will help to pave the way not only for necessary geographic expansion, but for the discovery of opportunities like network-sharing.

### 2.3.2 Future Market Development

Based on information from GSM Association (shown below in Figure 2.1), Western European GSM use is expected to comprise approximately 49% of world GSM cellular service in 2005. What is evident in the figures below is that GSM's role in global mobile cellular market is expected to decline, not only in terms of tapering subscriber numbers, but vis-à-vis giving way to other more advanced systems like GPRS – and ultimately bowing to IMT-2000 (UMTS in Europe). Based on forecasted data in Europe, it is likely that the number of UMTS subscribers will surpass the number of GPRS subscribers in 2004, and then go on to surpass GSM subscribers in Western Europe just after that by the end of the year (See Figure 2.3). This will occur just as GSM, GPRS, and UMTS subscribership, according to operators, is evening out across available systems (See Figure 2.2). Based on the assumption that the 'spread' of mobile users are in majority comprised of these GSM, GPRS, UMTS, and HSCSD systems, it appears that the total number of mobile subscribers in Europe will be somewhere around 697 million by 2005, about 40% of which will be UMTS users.

**Figure 2.1: Forecasted Adoption of GSM Mobile Phones in Western Europe and the World**

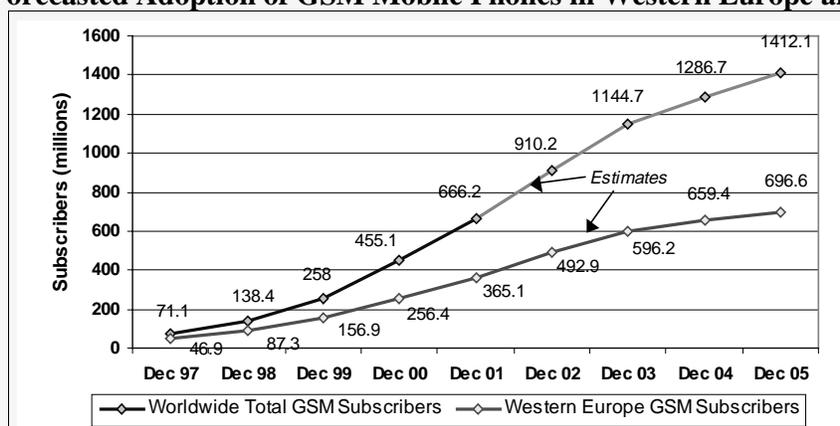

*Source: www.gsmworld.com*

---

[36] "GSM Association Subscriber Statistics". The GSM Association (accessed October 2001) Link: http://www.gsmworld.com/membership/ass_sub_stats.html.

**Figure 2.2: Comparison of 2G / 2.5G / 3G subscribership in Europe**

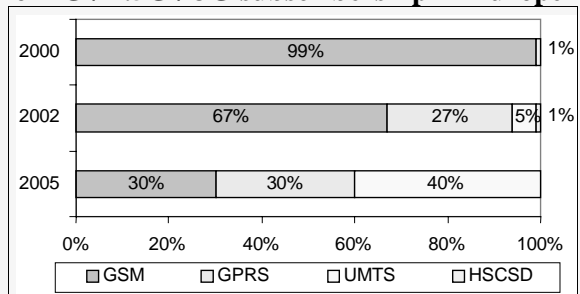

In response to a survey question:  How will your subscribers break down by network technology?
(Averages from 22 operators responding)

Source: *Forrester Research*

**Figure 2.3: Forecasted Subscribers for GSM, GPRS, UMTS and HSCSD Systems in Europe**

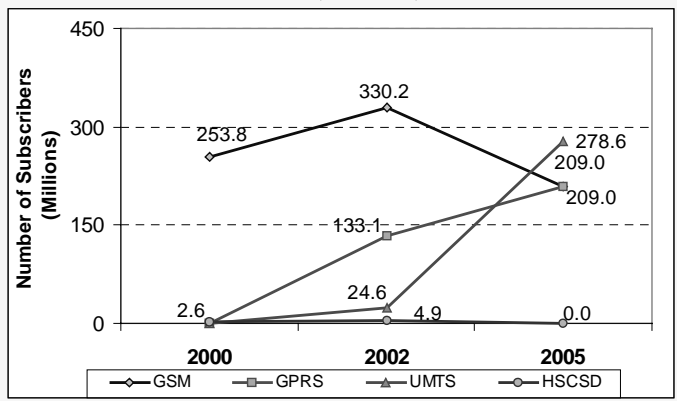

Source: *ITU analysis based on Forrester and GSMWorld.com research*

## 2.4   Licensing GSM

Throughout the 1980s, national governments were more often than not free to choose licenses, and with the exception of the UK, issued the first GSM licenses to their national PTT's.  "Public telephone operators (PTOs) in five European Community member states were given the opportunity to establish a strong presence in digital cellular GSM services long before their respective governments licensed second operators;  Belgacom, PTT Telecom Netherlands, Sip of Italy, Spain's Telefonica and Telecom Eireann all had head starts of a year or more on their competitors-to-be."[37]  Since the award of their first GSM licenses, many countries began to liberalize their telecommunications markets, usually introducing competition in the mobile wireless sector first.  Countries received numerous applications for their second GSM licenses, making the decision process more difficult than previous assignments.  Many countries began to add a financial bid to the list of selection criteria for their second digital license, while other countries continued with traditional comparative methods.[38]

The case of Spain, among others, is particularly interesting in this context.  After waiving the fee requirement for monopolist Telefonica's GSM license (Spain's first), the Spanish government subsequently saw "nothing wrong with its requirement to tender for a second private mobile telecomms license, and [to] request companies [to] make a payment to the treasury"[39]. (See Table 2.2 for further similar examples).

---

[37] "*PTTs steal a lead as GSM competition progress slows*".  Mobile Communications, <u>Financial Times Business Reports Technology File</u>, June 17, 1993, p.3.

[38] Spicer, Martin.  "*International Survey of Spectrum Assignment for Cellular and PCS*".  <u>Wireless Telecommunications Bureau,</u> Federal Communications Commission, September 1996. (accessed October 2001) Link: http://www.fcc.gov/wtb/auctions/data/papersAndStudies/spicer.html.

**Table 2.2: Digital License Assignment Patterns**

| First Digital License Assigned to PTT/Wireline Carrier |
| --- |
| Australia, Austria, Belgium, Ireland, Italy, Korea, France, Germany, HK, Spain, Sweden |
| **Countries using Financial bids for Second Digital License** |
| Australia, Austria, Belgium, Ireland, Italy, New Zealand, Poland, Spain |
| **Countries NOT using Financial Bids for Second Digital Licenses** |
| France (& PCN), Germany (&PCN), HK (& more), Korea, Sweden, UK |

*Source*: The Wireless Telecommunications Bureau. link: http://www.fcc.gov/wtb/auctions/papers/spicer.html

Governments in general could not be deprived of their individual gain from implementing the new GSM standard and from participating in the network. While there certainly existed incentives for governments to support their own national champion's quest for scale economies in equipment markets by seeking the adoption of a standard advocated by their own domestic manufacturer, the logic of networks combined with national telecom monopolies ensured that no cooperating party had to fear vastly unequal returns – even in case a national corporate champion lost out in the initial fight over the specifics of the network standard.[40]

By 1992, Finland (12/91), Germany (6/92), Denmark (7/92), France (7/92) , the United Kingdom (7/92), Sweden (9/92), Italy (10/92), and Portugal (10/92) were among the first countries in the world to launch their GSM services.

### 2.4.1    GSM Radio Spectrum

The ITU, which manages the international allocation of radio spectrum, allocated the 890-915 MHz bands for the uplink (mobile station to base station) and 935-960 MHz bands for the downlink (base station to mobile station) for mobile networks in Europe. "…Since this range was already being used in the early 1980s by the analog systems of the day, the CEPT had the foresight to reserve the top 10 MHz of each band for the GSM network that was still being developed."[41] It should be noted that the World Radio-Communications Conference (WRC) in 1992 identified frequency bands for FPLMTS (Future Public Land Mobile Telecommunications Systems), which is in fact the original name of IMT-2000 (UMTS).[42] The existing second-generation bands for second-generation GSM services consist of spectrum between 862 and 960 MHz and the totality of the GSM1800 band 1710 - 1880 MHz.

## 3    A Look Ahead at IMT-2000

## 3.1    From GSM to IMT-2000

The relationship between 2G and 3G is captured intrinsically in the migration process. The migration to 3G-services from 2nd generation systems is a broad topic area, depending on the starting point of the analysis; for example, CDMA-based systems have a very different road to IMT-2000 than TDMA counterparts. Such systems point to 'CDMA 2000' systems as equivalent to '3G', while for TDMA systems (including GSM), the Ericsson-proposed W-CDMA standard represents attainment of '3G'. Interestingly, CDMA-based carriers believe that their migration path[43] will be more inexpensive than that of GSM/TDMA-based carriers, because many will have only to change channel cards in the base stations and upgrade the network software as opposed to implementing entire network overlays. In any case, the focus of the present analysis will remain the path of GSM towards 3G.

---

[39] According to the Spanish press, two bids were received by the Spanish Ministry of Posts and Telecommunications for the second GSM network license. The Airtel consortium is understood to have bid Pta 85 billion, while the Cometa SRM group bid a conditional Pta 89 billion. "*Spain - Furor Over Cellular Telephony Licensing*". Newsbytes News Network. December 21, 1994.

[40] Bach, p.8.

[41] Scourias, John. "*Overview of the Global System for Mobile Communications*". University of Waterloo, 1997. (accessed July 2001) Link: http://www.shoshin.uwaterloo.ca/ ~jscouria /GSM/gsmreport.html.

[42] "*Communication From the Commission to the European Parliament and the Council*", The World Radiocommunications Conference 1997, (WRC-97). The European Commission. Brussels, [COM(97) 304 final]. (accessed October 2001) Link: http://europa.eu.int/ISPO/infosoc/legreg/docs/97304.html#Heading8.

Enhancements upon 2nd generation GSM systems include HSCSD (High Speed Circuit Switched Data), GPRS (General Pack Radio Service), and EDGE (Enhanced Data Rate for GSM evolution) – all of which allow for higher data transmission rates. (See Figure 3.1 and Table 3.2) The goal of GSM migration is to reach UMTS, which is part of the ITU's 'IMT-2000' vision of a global family of 'third-generation' (3G) mobile communications systems. All of these 2.5 generation systems are now well on their way to development and deployment – and the question now is which one will be most relevant, versatile, cost-effective, and able to cope with the demands of a complex telecommunications service landscape? Which system will succeed in effectively offering which services? (See Table 3.1) And, will these 'half-steps' toward elusive 3G-roll-out pre-empt the need for 3G itself, or just delay its introduction?[44]

**Table 3.1: Comparative View on Services/Applications**

| Period | Major Technology Introduction | New Internal/External Applications |
|---|---|---|
| **Up to 2000** | 2 G | • Telephone<br>• Email<br>• SMS<br>• Digital Text Delivery |
| **2001 to 2002** | 2.5 G | • Mobile Banking<br>• Voicemail, Web<br>• Mobile Audio Player<br>• Digital Newspaper Publishing<br>• Digital Audio Delivery<br>• Mobile Radio, Karaoke<br>• Push Marketing/ Targeted programs<br>• Location-based services<br>• Mobile coupons |
| **2003 and beyond** | 3 G | • Mobile videoconferencing<br>• Video Phone/Mail<br>• Remote Medical Diagnosis and Education<br>• Mobile TV/Video Player<br>• Advanced Car Navigation/ City Guides<br>• Digital Catalog Shopping<br>• Digital Audio/Video Delivery<br>• Collaborative B2B Applications |

*Source*: International Telecommunication Union

---

[43] CDMA systems can use a data-only 2.5G standard, called High Data Rate (HDR), capable of data rates up to 1.4 Mbit/s delivered to mobile wireless data customers in a fixed mode. 1XRTT, an advanced version of IS-95 for mobile users, delivers transmission speeds up to 144 Kbit/s. "*The future of 3G*", <u>EDN</u>, Boston. June 7, 2001, p. S9.

[44] It is interesting to note in light of this question that "…96% of operators in an interview by Forrester believe that 3G (or UMTS) is important to their business plans." Godell, Lars. "*Europe's UMTS Meltdown*". <u>Forrester Research Report</u>, December 2000, p.3.

**Figure 3.1:  A Step-by-Step Towards IMT-2000 (UMTS)**

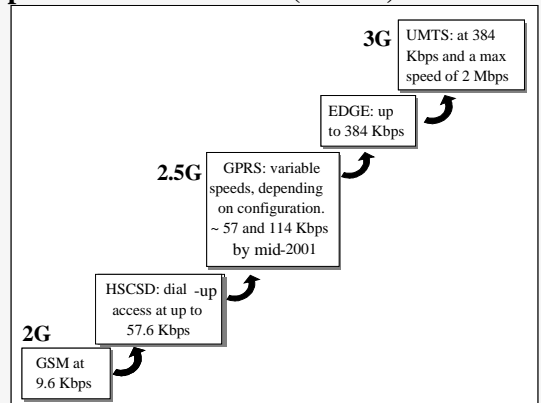

Note: This is an illustrative figure only.  Please note that a shift toward additional spectrum occurs after the EDGE component, upon the 'leap' to UMTS.  There is some debate about the status of 'EDGE' as potential equivalent of UMTS / IMT-2000, given that its data transmission capacity is close to expected 3G rates (UWC-136 or EDGE is recognized under the ITU's IMT-2000 umbrella); on the other hand, it appears that there may be diminishing scope for the deployment of EDGE in future.
*Source:  International Telecommunication Union*

It is interesting to mention that some feel the jump from 2G to 2.5G will be more dramatic than that from 2.5G to 3G.  "…the big job for the operator is not going from GPRS to UMTS, it's actually going from GSM into GPRS, because you change completely the business model, going from time-based to volume-based charging.  You also go from more traditional-type services to more internet-based services."[45]  Although 2.5G technologies were expected to smooth the transition to 3G, WAP experiences proved to be less than satisfactory.  To a certain extent, WAP showed the effect of excessively high expectations on technologies in markets when they under-perform.

It is unlikely that there will be a sudden jump from today's GSM networks to the 3G networks of tomorrow. GSM-based services, as mentioned before, rely on digital transmission between base stations and handsets with high-speed connections to and from the centers equipped with circuit switches.    At 9.6 Kbit/s, transmission is slow, and the architecture itself is unsuitable for data traffic or streaming as it is circuit-switched rather than packet-switched.   While GPRS seems to be an obvious migration step for GSM operators, next steps require further evaluation.  It is also important to note that through the course of the transition, it is not necessarily the case that the early 3G networks – when they appear – will be packet-switched from their debut; this evolution to packet-based networks is likely to occur over some time as systems are tested and proven.

**Table 3.2:  Detailed Comparison of 1st, 2nd, and 3rd Generation Technologies**

|  |  | Technology | Bandwidth (Kbit/s) | Features |
|---|---|---|---|---|
| **First Generation Mobile** | AMPS/ NMT | Advanced Mobile Phone System  Nordic Mobile Telephony | 9.6 | • Analog voice service • No data capabilities |
| **Second Generation Mobile** | GSM | Global System for Mobile Communication | 9.6 → 14.4 | • Digital voice service • Advanced messaging • Global roaming • Circuit-switched data |
|  | HSCSD | High-Speed Circuit Switched Data | 9.6 → 57.6 | • Extension of GSM • Higher data speeds |

---

[45] "*Bridge Over Troubled Water*", <u>Mobile Matters</u>,  May 2001,  p.56.

| | | | | |
|---|---|---|---|---|
| | GPRS | General Packet Radio Service | 9.6 → 115 | • Extension of GSM<br>• Always-on connectivity<br>• Packet-switched data |
| | EDGE | Enhanced Data Rate for GSM Evolution | 64 → 384 | • Extension of GSM<br>• Always-on connectivity<br>• Faster than GPRS |
| **Third Generation Mobile** | IMT-2000/UMTS | International Mobile Telecommunications 2000 / Universal Mobile Telecommunications System | 64 → 2,048 | • Always-on connectivity<br>• Global roaming<br>• IP-enabled |

*Source*: Forrester Research

### 3.1.1 HSCSD (High-Speed Circuit Switched Data)

HSCSD is a natural evolution of the existing circuit-switched data capability of traditional 2G GSM networks. With today's GSM network standards, it is already possible to transmit narrowband data and digital fax over the TDMA air interface. The methodology is akin to setting up a GSM voice call or perhaps to making a connection over a fixed line PSTN with the use of a modem. The user establishes a connection (or circuit) for the whole duration of that communication session. To set up the circuit, a call set-up process is involved when dialling the called party; network resources are allocated along the path to the end destination.

Within the existing GSM encoding techniques, the maximum circuit-switched data (CSD) speed is 9.6 Kbit/s or with improved encoding, up to 14.4 Kbit/s. The GSM TDMA interfaces can assign up to 8 time division slots per user frequency, not all of which are always used. Typically one is allocated for voice, while other slots may be allocated for fax and data. The availability of these time slots makes it possible to expand the existing CSD into HSCSD. The transition to HSCSD is not a difficult one for an existing 2G operator, and typically only necessitates a software upgrade of the Base Stations Systems (BSS) and Network and Switching System (NSS) systems.

A potential technical difficulty with HSCSD arises because in a multi-timeslot environment, dynamic call transfer between different cells on a mobile network (called 'handover') is complicated, unless the same slots are available end-to-end throughout the duration of the circuit switched data call. The second issue is that circuit switching in general is not efficient for bursty data/Internet traffic. The allocation of more circuits for data calls, with typically longer 'hold' times than for voice calls, creates the same problems that fixed line PSTN operators have experienced with the tremendous growth of Internet traffic – i.e., too few resources in their circuit switched networks.[46]

### 3.1.2 GPRS (General Packet Radio Service)

"GPRS is seen as a closer step towards UMTS and… with increased data speeds – will sit somewhere in between 2G and 3G rates – it will introduce a more functional medium in which consumers will see the potential of 3G."[47] GPRS is an overlay technology that is added on top of existing GSM systems. In other words, the GSM part still handles voice, and handsets are capable of supporting both voice and data (via the overlay) functions. GPRS essentially supplements present-day circuit-switched data and short message services (SMS), and serves as an enabler of mobile wireless data services, and an optimizer of the radio interface for bursty packet mode traffic. The upgrade to GPRS is easy and cost effective for operators, as only a few nodes need to be added. According to the Dec 1998/January 1999 issue of Mobile Communications International, "…the move to GPRS will be worth the expense because it will position operators well for 3G. Once carriers have built a packet subsystem for GPRS, they will be able to add additional 3G services as needed through co-sited GSM and WCDMA base station subsystems."[48]

GPRS is packet-based and promises data rates from 56 up to 114 Kbit/s, as well as continuous connection to the Internet for mobile phone and computer users. More specifically, packet-switching means that GPRS radio resources are used only when users are actually sending or receiving data; available radio resources can be concurrently shared between several users. This efficient use of scarce radio resources means that large numbers of GPRS users can potentially share the same bandwidth and be served from a single cell. The actual number of users supported depends on the application being used and how much data is being transferred. Because of the spectrum efficiency of GPRS, there is less need to build in idle capacity that is only used in peak hours. GPRS therefore lets network operators maximize the use of their network resources in a dynamic and flexible way, along with user access to resources and revenues.

---

[46] "*Wireless Overview For Non Wireless Professionals*". White Paper by <u>Nortel Networks</u>. (accessed July 2001) Link: www.nortel.com.

[47] "*Bridge Over Troubled Water*", <u>Mobile Matters</u>, May 2001, p.56.

[48] "*Scheduling a Date with Data*", <u>Mobile Communications International</u>, December 1998/January 1999.

GPRS is essentially based on "regular" GSM (with the same modulation) and is designed to complement existing services of such circuit-switched cellular phone connections such as SMS or cell broadcast. GPRS should improve the peak time capacity of a GSM network since it simultaneously transports traffic that was previously sent using CSD through the GPRS overlay, and reduces SMS Center and signalling channel loading. In theory, GPRS packet-based service should cost users less than circuit-switched services since communication channels are being used on a shared-use, as-packets-are-needed basis rather than dedicated only to one user at a time. It should also be easier to make applications available to mobile users, and WAP or i-mode should be far more attractive for the user. In addition to the Internet Protocol, GPRS supports X.25, a packet-based protocol that is used mainly in Europe.

GPRS for the time being has fallen short of theoretical 171.2 Kbit/s maximum speed, one reason being the technical limitations of currently available handsets. Nevertheless, GPRS rollouts are expected to help counterbalance previous disappointments associated with WAP-based services/technology; hope is not lost, particularly according to the Gartner Group, that WAP can be a primary driver for mobile data revenue growth in the next three to five years. GPRS has the potential to 'help WAP get back on its feet again', according to John Hoffman of the GSM Association.[49]

### 3.1.3    EDGE, Enhanced Data GSM Environment

Enhanced Data rates for Global Evolution (EDGE) is a radio based high-speed mobile data standard that allows data transmission speeds of 384 Kbit/s to be achieved when all eight timeslots are used. EDGE was formerly called GSM384, and is also recognized as 'UWC-136' under the ITU's specifications for IMT-2000. It was initially developed for mobile network operators who failed to win spectrum for third generation networks, and is a cost-efficient way of migrating to full-blown 3G services. It gives incumbent GSM operators the opportunity to offer data services at speeds that are near to those available on UMTS networks.

EDGE does not change much of the core network, however, which still uses GPRS/GSM. Rather, it concentrates on improving the capacity and efficiency over the air interface by introducing a more advanced coding scheme where every time slot can transport more data. In addition, it adapts this coding to the current conditions, which means that the speed will be higher when the radio reception is good. Implementation of EDGE by network operators has been designed to be simple, with only the addition of one extra EDGE transceiver unit to each cell. With most vendors, it is envisaged that software upgrades to the BSCs and Base Stations can be carried out remotely. The new EDGE capable transceiver can also handle standard GSM traffic and automatically switches to EDGE mode when needed. 'EDGE-capable' terminals are also needed, since existing GSM terminals do not support new modulation techniques, and need to be upgraded to use EDGE network functionality.

EDGE can provide an evolutionary migration path from GPRS to UMTS by more expeditiously implementing the changes in modulation that are necessary for implementing UMTS later. The main idea behind EDGE is to squeeze out even higher data rates on the current 200 kHz GSM radio carrier, by changing the type of modulation used, whilst still working with current circuit (and packet) switches.

In addition, the TDMA industry association, the "Universal Wireless Communications Corporation", has introduced what it calls EDGE Compact. This is an even more spectrum-efficient version of EDGE that will support the 384 Kbit/s mandated packet data rates, whilst requiring only minimum spectral clearing. In fact, as a result of this, EDGE has been renamed Enhanced Data Rates for GSM and TDMA Evolution. EDGE is planned to be commercially available end of year 2001.[50]

---

[49] "*Bridge Over Troubled Water*", <u>Mobile Matters</u>, May 2001.  p.56.
[50]" *Comprehensive information about GPRS and Edge*".    (accessed July 2001)  Link:  http://www.3ggeneration.com/gprs_and _edge.htm.

When describing the services to which 3G technologies aspire, it is crucial to bear in mind that there is a difference between what is possible in reality and what is 'hype' vis-à-vis data speeds. That said, however, any reference to 'hype', is by definition a reference to the expectations of 3G created largely from the press and other sources less likely to have significant technical mastery of the respective systems. The ITU from the early phases of IMT-2000 development, has given unambiguous recommendations for the exact testing conditions under which various technical specifications for systems have been developed. How these recommendations have been commonly translated for the mass market, however, has resulted in somewhat 'less-than-scientific' evaluations, which in turn has contributed to the afore-mentioned 'hype'.

Nonetheless, it is an interesting exercise to compare the 'hyped' market expectations with 'reality'. In practice, data throughput is inversely proportional to the sizes of the cells that are covered by one transceiver (base station). The higher the data throughput sought, the smaller and more numerous the cells deployed by operators, and hence the greater the difficulties in reaching very rural areas. Figure 3.2 illustrates the divergence (in the European case, specifically) between 'hype' and reality, by laying out the deployment of the various migration steps towards UMTS. It appears that relative to the 3G 'hype', 'city' 3G deployment is unlikely to be realized before 2003, although launch dates have been set optimistically for 2002.

**Figure 3.2: From GSM to UMTS: Likely Paths to 3G**[51]

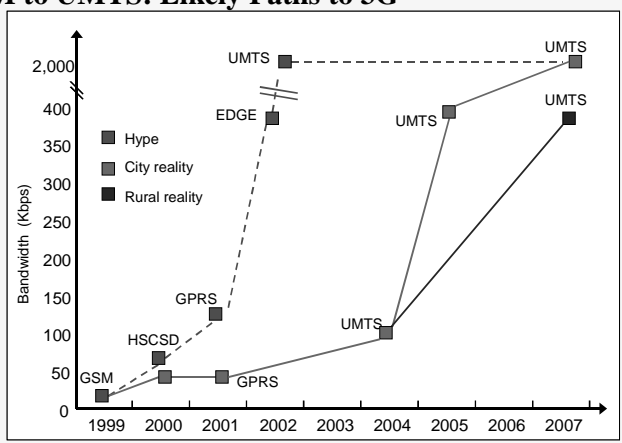

Source: *Forrester Research*

Operators running GSM 1800 networks will have an advantage over those running GSM 900 networks because the higher frequency and lower power are closer to providing good coverage at UMTS frequencies. Many GSM 1800 cell sites will be re-usable. From GSM, at 9.6 Kbit/s, to EDGE and UMTS, at 384 Kbit/s, the percentage increase in data throughput is less than the figures suggest. Nevertheless, the faster speeds are sufficient for applications such as e-mail, Short Message Service (SMS) and access to the Internet and corporate intranets. Network operators will not able to guarantee customers maximum data throughput at any instant during a call session. Mobile packet networks are headed for "best effort" service, at least for another four to five years; high-quality services based on UMTS must bear the caveat that data transmission may reach anywhere 'up to 2 Mbit/s'. It appears that nothing is guaranteed.

In terms of the migration from 2G to 3G services, over half of the operators in a recent survey by ARC Group believed that GSM operators in their country would adopt GPRS, while only a quarter expected that EDGE technology would be deployed.[52] By 2002, 65% of those surveyed said that commercial consumer 3G services would be up and running in their country, with 42% predicting an initial data transmission speed of over 90Kbit/s, some way short of the maximum 2Mbit/s expected to be available from UMTS.[53] As was the case with the introduction of GSM in the 1980s, important regulatory issues (e.g. licensing, numbering, and frequency band allocation) for UMTS in Europe have been addressed in order to create the optimal conditions for investment and a predictable environment for the emergence of alliances that can develop it.

---

[51] "*Mobile's High Speed Hurdles*", Forrester Research Report, March 2000.

[52] "*Operators Express Concern Over Handsets*" Arc Group, January 16, 2001. (accessed August 2001) Link: http://www.arcgroup.com/press2/ cut_concernhandsets.htm.

[53] Ibid, (accessed August 2001) Link: http://www.arcgroup.com/press2/ cut_concernhandsets.htm.

## 3.2    IMT-2000 Technology

The vision of IMT-2000 (3G) networks is defined by a single standard comprised of a 'family of technologies' intended to provide users with the ability to communicate anywhere, at any time, with anyone. 3G network architecture is based on two main principles:  one is that mobile cellular networks should be structured to maximize network capacity, and the other is to offer multimedia services independently of the place of the end users.   The 3G umbrella encompasses a range of competing mobile wireless technologies, namely CDMA-2000 and WCDMA.

European UMTS (which stands for Universal Mobile Telecommunications System), falls within the ITU's IMT-2000 vision of a global family of 3G mobile communication systems.  It includes WCDMA radio access technologies, together with a core network specification based on the GSM/MAP (Mobile Application Part) standard.   As reflective of 3G in Europe and specifically the focus of this paper in GSM context, UMTS is actually intended to provide the kinds of data speeds and protocols to allow people with appropriate handsets to access the Internet, watch movies, exchange large data files and have video conference calls to and from locations of temporary choice and convenience.  The new network, improving upon previously described shortcomings, has to allow for data traffic, which comes in unpredictable bursts, voice conversations, which should not be interrupted, and the streaming of large contents like movies.  The goal for 3G is to provide standard facilities good enough for mobile devices to handle color video.

3G communications are based on standards that are intended to ensure global interoperability and standardized usage of spectrum frequency.  Across Europe, countries have adopted different policies for allowing the development of 3G services: some, like Germany and the United Kingdom, auctioned the rights to use the designated spectrum; others, like Finland and Spain, invited applicants and selected providers for various features and promises; and others, like Sweden, are sharing the risk by charging a royalty on future 3G revenue.  This is discussed further in Section 3.6.

IMT-2000 itself offers the capability of providing value-added services and applications on the basis of a single standard.   The system envisages a platform for distributing converged fixed, mobile, voice, data, Internet and multimedia services.   One of the key aspects of its vision is the provision of seamless global roaming, enabling users to move across borders while using the same number and handset.  It also aims to provide seamless delivery of services, over a number of media (including satellite, fixed, etc.).  It is expected that IMT-2000 will provide higher transmission rates than currently possible, i.e., a minimum speed of 2Mbit/s for stationary or walking users, and 348 Kbit/s in a moving vehicle.

## 3.3    The History of IMT-2000

In the mid-1980's, the ITU created this 'single standard' entitled 'IMT-2000', "International Mobile Telecommunications", to serve as the base of the third generation system for mobile communications. Needless to say, the ambiguous definition ascribed to this 3G standard plainly reflects the fact that challenges arose to the goals of interoperability and seamless integration for the eventual provision of global roaming service.  In 2000, unanimous approval was given for the technical specifications for third generation systems under this same brand name.  IMT-2000 is thus the result of collaboration of many entities, both inside and outside the ITU (ITU-R and ITU-T, and 3GPP, 3GPP2, UWCC, etc.) – although the extent to which it can be said to have been reflective of successful cooperation is questionable.    It seems rather that the essence of IMT-2000 is captured by serious compromise, where the costs of cooperation have been out shadowed by the benefits of protecting existing legacy investments.

It is obvious that proponents of the different approaches to 3G technologies – CDMA2000 (US, Korea), and W-CDMA (Europe, Japan) were not able to agree on a single standard – hence the variety of 'flavours' of wideband CDMA that comprises achievement of "3rd generation" status.  IMT-2000 therefore, as mentioned earlier, consists of a 'single standard of a family of technologies', which implies the need for multiple mode and multiple band handsets capable of handling various optional mode and frequency bands.  The system as a whole is said to be highly flexible, capable of supporting a wide range of services and applications.   This is one way of looking at a standard that did little to actually standardize the international and interoperable advancement to the 3rd generation.

The ITU has clearly indicated that at the heart of the IMT-2000 project is the objective to raise awareness of the importance and reach of IMT-2000 as a global, harmonized mobile personal communication system and access platform, with emphasis on its role in the deployment of the global wireless information society. While the ITU is committed to its role as the 'best-positioned' organization to act as facilitator and coordinator of global standards development, global frequency spectrum harmonization, and global circulation of IMT-2000 terminals, the compromise reflected in the formation of IMT-2000 casts this crucial role in an uncertain light. One of the goals of IMT-2000 is to provide an evolutionary path from 2G systems to 3G systems and to protect existing investments in legacy 2G systems. But this seems to come at the cost of allowing for tremendous uncertainty and considerable market fragmentation, whereas, in the case of GSM, there was a far clearer incentive to keep the notions of international roaming and the 'collective good' at the forefront of the agenda.

According Dr Bernd Eylert, Chairman of the UMTS Forum, "…the market has already demonstrated the attraction of global standards operating in harmonized spectrum plans - as seen in the past successes of AMPS and GSM - and by adopting the same radio planning methodology as other ITU regions. Furthermore, operators using open standards in harmonized spectrum have the opportunity to compete on service, coverage, quality and price... and this will always benefit the end user."[54]  This point of view implies that 'open standards' and enhanced 'flexibility' in the diversity of the IMT-2000 umbrella is actually the key to ultimately best servicing the consumer.

The UMTS Forum, an international, non-profit, independent body created in 1996 (based in the U.K.), is among these entities involved in standardization and committed to the successful introduction and development of UMTS/IMT-2000, through the creation of cross-industry consensus. It currently has 250 member organisations drawn from the mobile operator, regulatory, supplier, consultant, IT and media/content communities, and works on issues like technical standards, spectrum, market demand, business opportunities, terminal equipment circulation and convergence between the mobile communications and computing industries.[55]  The American counterpart of the UMTS Forum is the CDMA Development Group (CDG), an international consortium based in the U.S., and comprised of leading CDMA service providers and manufacturers, who have joined together to lead the adoption and evolution of CDMA wireless systems around the world. The CDG is working to ensure interoperability among systems, while expediting the availability of CDMA technology to consumers.[56]

The IMT-2000 standard accommodates five possible radio interfaces (or flavours) based on three access technologies (FDMA, TDMA, and CDMA). (See Figure 3.3) The two main interfaces fall under the 'Wideband–CDMA', and the US-supported 'cdma2000' categories. The W-CDMA standard includes the European usage of W-CDMA (generally recognized in the form of UMTS), and the Japanese standard used by NTT DoCoMo. Cdma-2000 is a Telecommunications Industry Association (TIA) standard for third-generation technology, that is an evolutionary outgrowth of cdmaOne from the United States. Both W-CDMA and cdma2000 are mainly based on 'Frequency-Division-Duplex' (FDD) frameworks. A third interface falls under the TD-SCDMA category, the radio interface proposed by China and approved by the ITU, which is based on 'Time-Division-Duplex' (TDD)).[57]  The fourth interface falls under the TDMA category (UWC-136 ('Universal Wireless Communications'-136)), which is also otherwise known as EDGE; this was developed by CDMA AMPS operators, many of which have since developed different migration strategies. Finally, the last interface falls under the FD-TDMA category (known as DECT+ for use in Europe), which performs like IMT-2000, but is in fact used mainly for indoor environments. Essentially, it is evident that of the five main 'flavours' depicted in Figure 3.3, three are the most prominent in terms of applicability and future potential.

---

[54] "*Brazil is Poised to Embrace Global 3G IMT-2000 Opportunity, Says UMTS Forum*". UMTS Forum, March 31,2000. (accessed August 2001) Link: http://www.umts-forum.org/press/article034.html.

[55] The UMTS Forum. (accessed October 2001) Link: http://www.umts-forum.org/.

[56] The CDMA Development Group. (accessed October 2001) Link: http://www.cdg.org.

[57] It should be noted that TDD is also used for UMTS (which is in fact a combination of the components of W-CDMA and TD-SCDMA solutions); UMTS more aptly fits under the category of IMT-TC.

**Figure 3.3:  IMT-2000 Terrestrial Radio Interfaces**

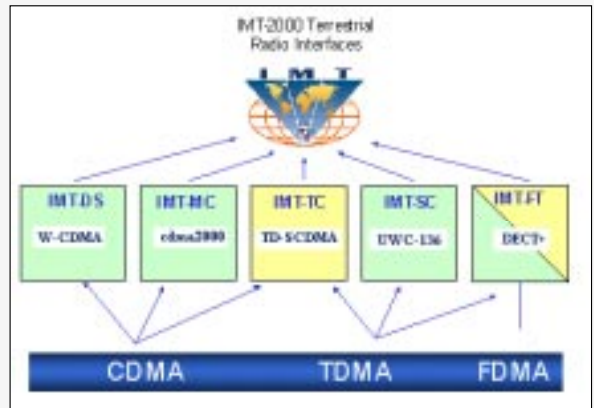

*Value-added services and worldwide applications development on the basis of one single standard accommodating five possible radio interfaces based on three technologies*

Source: International Telecommunication Union

## 3.4    Laying the Groundwork for 3G Success

In order for IMT-2000 to be possible, it has been necessary over the past few years to create the impetus for its realization; part of this, of course, being driven by the simple fact that existing 2G circuit-switched systems will be inadequate for forthcoming data transmissions.  "…Existing systems like GSM are running out of capacity… and the mobile phone market is growing at an annual rate of about 55% … it has been estimated that 80% of the population in the European Union will have some form of mobile communicator by the year 2020…"[58]  Given these figures, some major efforts were undertaken in 2000 to make a 'first step' toward 3G.

### 3.4.1    Addressing the Need for 3G Spectrum Expansion

The WRC 2000, the international forum which serves to provide the technical, operational and regulatory conditions for the use of radio frequency spectrum and satellite orbits, was critically important in its management of radio frequency spectrum for 3rd generation technologies.  The awareness that more spectrum would be needed was at the forefront of the WRC's mission.  And it provided, in what can now be considered as a landmark decision, the conditions under which the industry could continue to develop and deploy a host of sophisticated new radio-based communications systems over the next few years.[59]  With 3G mobile systems due to come into service very soon in several countries, it was imperative that an increase in available spectrum be ensured for 3G services.

The existing spectrum identified back in 1992 for GSM upon which licensing is now taking place around the world, was based on a model in which voice services were considered to be the major source of traffic, and only low data rate services were considered.  In fact, Resolution 223 adopted at WRC-2000 found that ITU studies demonstrated the need for approximately 160 MHz of spectrum in addition to that identified at WRC-92, and in addition to the spectrum already being used for first and second generation wireless services.[60]  The need for added spectrum stemmed from three main considerations: the first being that the number of users is expected to reach an estimated 2 billion worldwide by 2010[61], the second being the rapid growth of mobile data services, mobile e-commerce, wireless internet access and mobile video-based services, and the third being the need to secure common spectrum worldwide for global roaming and cheaper handsets.

---

[58]  Berg, Andreas, "*UMTS – Universal Mobile Telecommunications System*". Helsinki University of Technology (accessed October 2001) Link: http://www.tml.hut.fi/Opinnot/Tik-111.350/1998/esitykset/Umts/UMTS.html.

[59]  "*Thumbs up for IMT-2000*",  International Telecommunication Union Press Release, May 30, 2000.  (accessed October 2001) Link: http://www.itu.int/ newsarchive/press/releases/2000/12.html.

[60]  Provisional Final Acts of the World Radiocommunication Conference  (Istanbul, World Radiocommunication Conference 2000), Resolution 223, § h.

[61]  "*The UMTS Forum – Shaping the Mobile Future*", The UMTS Forum, October 2000. p.3. (accessed October 2001) Link: http://www.umts-forum.org/brochures/UMTS.pdf.

All of the spectrum between 400 MHz and 3 GHz is technically suitable for third generation mobile. The entire telecommunication industry, including both private sector and national and regional standards-setting bodies gave a concerted effort to avoid the fragmentation that had thus far characterized the mobile market. WRC approval meant that for the first time, full interoperability and inter-working of mobile systems could be achieved. Three common bands are available on a global basis for countries wishing to implement the terrestrial component of IMT-2000. The three bands identified for use by IMT-2000 include one below 1 GHz, another at 1.7 GHz (where most of the second-generation systems currently operate to facilitate the evolution, over time, of these systems to third generation), and a third band in the 2.5 GHz range. These complement the band in the 2 GHz range already identified for IMT-2000. The Conference also identified the use of additional frequency bands for the satellite component of IMT-2000. For the European UMTS (3G) network specifically, bands are available in a 155 MHz wide spectrum in the 1.9 and 2.1 GHz band.[62]

The agreement provides for a high degree of flexibility to allow operators to evolve towards IMT-2000 according to market and other national considerations, and gives a green light to the mobile industry worldwide in confidently deploying IMT-2000 networks and services. Making use of existing mobile and mobile-satellite frequency allocations, it does not preclude the use of these bands for other types of applications or by other services to which these bands are allocated – a key factor that enabled the consensus to be reached. While the decision of the Conference globally provides for the immediate licensing and manufacturing of IMT-2000 in the common bands, each country decides on the timing of availability at the national level according to individual need. This flexibility will also enable countries to select those parts of the bands where sharing with existing services is most suitable, taking account of existing licences.[63]

## 3.5   The 3G Market

As the path to UMTS in particular is inextricably linked to the history of GSM, it is interesting to look at what factors were driven by GSM penetration, and how they have impacted the forecasts and general market conditions for 3$^{rd}$ generation mobile technologies. Key drivers and justification for the exorbitant sums spent on 3G spectrum licenses lie partly in aspects such as those featured below.

One way of gauging the likelihood of 3G's success is to look at one of its closest forerunners: SMS via GSM. Some consider it to be the best indicator of the money-generating potential of the mobile internet, assuming that SMS usage can be easily translated to demand for data on mobile devices. The widespread success of SMS in Western Europe contributed significantly to mobile data revenue in 1999 and showed that consumers will use mobile phones for more than just voice. Most importantly – in terms of its potential implications for IMT-2000 – it must be recalled that SMS was a value-added service innovation which could not have been predicted when the service was first launched in the 1990's.

The GSM Association estimated that GSM networks transported one billion messages worldwide in October 1999, and SMS revenue apparently comprised a significant portion of overall service revenue figures in more mature markets such as Finland and Norway. By December, volume was up to two billion, and by March 2000 it was over three billion. Some 50 billion text messages were sent worldwide in the first three months of 2001 alone; "some 25.3 billion SMS text messages were sent in the first twenty-seven days of June 2001."[64]   Gartner's Dataquest expects SMS usage and revenue to continue to grow strongly across Western Europe during the next two years, though Forrester actually sees a slight decline – from 8% to 7% in 2003 – as other forms of data traffic gain precedence on mobile networks. (See Figure 3.4)  Global income from text and messages in 2001 is expected to reach $18.9 billion on total mobile phone revenues of $400 billion, according to research group Ovum.

---

[62] "The spectrum is divided into 5 MHz carriers, but, since each carrier could be used either upstream or downstream, they are paired two by two (two times 5 MHz). The bandwidth in a cell depends on the size of the cell. The largest cells, called macro-cells, have a radius of about one kilometer and are limited to 114 Kbit/s. Smaller cells, called microcells, are as small as 400 meters in radius and can provide up to 384 Kbit/s. To provide higher-level data services an operator needs a third layer of even smaller cells, called pico-cells, with a radius of 75 meters. Only at this distance, and then only to almost stationary users, is it possible to provide 2 Mbit/s. In addition, the aforementioned bandwidths are shared by all users in the cell. If the total bandwidth is 384 Kbit/s in a cell, it can support 24 phone calls (at 16 Kbit/s) or six low-end video services (at 64 Kbit/s). It is unlikely that data services above 64 Kbit/s will be offered if a layer of pico-cells is not used. Operators that hold three paired carriers are the only ones that can build all three layers of cells and probably the only ones that will provide high-speed data services. The operators that are struck with only two paired carriers will not play in the same league." Montelius, Johan, "*GSM Subscribers to Carry Cost of UMTS License Fees*", Jupiter Media Metrix, September 18, 2000, p.1.

[63] "*World Radiocommunication Conference concludes on series of far-reaching agreements*". International Telecommunication Union Press Release, June 2, 2000. (accessed October 2001) Link: http://www.itu.int/newsarchive/press/releases/2000/13.html.

[64] Van Grinsven, Lucas. "*Mobile & Satellite: Nokia 3G guru cites SMS as key to wireless web success*". Reuters, June 28, 2001.

### Figure 3.4: Voice Traffic vs. Data Traffic Forecasting

How will operator revenues break down in 2000 and 2003?

|  | 2000 | 2003 |
|---|---|---|
| Voice Traffic | 90% | 68% |
| SMS | 8% | 7% |
| Mobile Internet Revenues | 3% | 25% |
| Data Traffic | 2% | 14% |
| Content Services | 0% | 4% |
| e -Commerce Commissions | 0% | 3% |
| Advertising Fees | 0% | 3% |

Source: *Forrester Research*

Data and Voice Revenue Forecasts, Western Europe (billion)

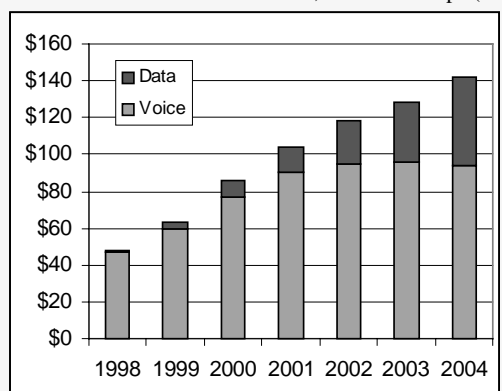

Source: *Gartner Dataquest*

Gartner Group expects that by 2004, mobile data in Western Europe will be a principal driver of increasing revenue, accounting for approximately 33% of mobile services revenue, up from 3% during 1999. (See Figure 3.4) By 2005 111 million customers – 63% of all subscribers – will access the mobile internet at least once monthly via push, pull and LBS, adding to carriers' top line revenues.[65] "Without doubt, data is becoming increasingly important for operators. Vodafone's D2, for example, derives 16% of revenue from mobile data and 11% of Sonera's mobile revenues are from SMS. Looking forward, Vodafone expects data service to account for more than 25% of revenues by 2004."[66]

**User Base Forecasts:**

Further evidence of strong forecasted market growth lies in the expanding mobile user base, as illustrated by the ITU below. As evident from analysis of the GSM market in the previous chapter, it is quite logical to assume that GSM development and growth has strongly influenced the extent of global cellular penetration. The number of worldwide mobile phone subscribers is predicted by some to reach 820 million by 2001, and there are likely to be more than a billion mobile users by 2003, and more than 2 billion in the next 10 years.[67] According to the ITU, at the start of the last decade there were just over 10 million mobile cellular telephone subscribers around the world, and this figure had grown by almost 70 times to over 725 million by the beginning of this year (2001).

Growth has been steady at an average of 50% per year since 1996. In Europe alone, mobile penetration exceeded 40% of the adult population at the end of 1999 — a figure that is likely to rise to 70% by 2005.[68] According to yet another source, the mobile penetration rate for Europe currently stands at approximately 60% and is forecast to grow to almost 80% by 2004.[69] At current growth rates, the number of mobile subscribers will surpass that of fixed telephones in the middle of this decade (see Figure 3.5). There are 35 markets – both developed and developing – where this transition has already taken place (see Table 3.3). In developing countries, competition and pre-paid cards are proving a powerful combination for driving mobile growth. The rise of mobile in developing countries in particular is perhaps most powerfully suggested by the fact that based on current growth, China will surpass the United States and emerge as the world's largest cellular market sometime this year (see Figure 3.6).

---

[65] McCarthy, Amanda. "*Mobile Internet Realities*". Forrester Research Report, May 2000.

[66] Bratton, William, Jameson, Justin, and Pentland, Stephen. "*Analysis: 3G madness – time for some moderation!*" TotalTele.com, July 16, 2001, p.2.

[67] "*The UMTS Forum – Shaping the Mobile Future*", The UMTS Forum, October 2000. p.3. (accessed October 2001) Link: http://www.umts-forum.org/brochures/UMTS.pdf.

[68] Butler, Andrew. "*Server Selection Strategies for WAP*", Gartner Group. June 13. 2000.

[69] "*European Overview*". Frost & Sullivan Research. 2001, p.2-1.

**Figure 3.5: Fixed and Mobile Lines, 'Big Picture' and 'Closer Up'**

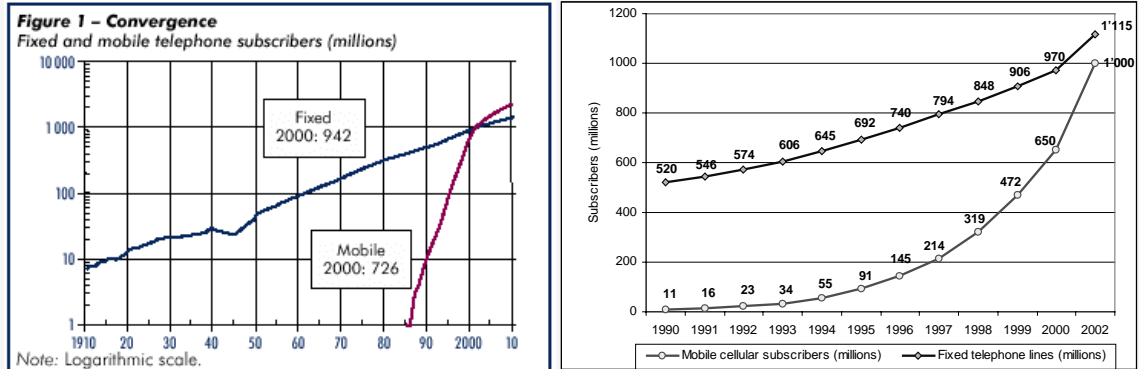

Source *International Telecommunication Union*

**Table 3.3: Economies Where Mobile Phones Have Overtaken Fixed Ones**

| 1993 | 1998 | 1999 | 2000 | |
|------|------|------|------|------|
| Cambodia | Finland | Austria | Bahrain | Philippines |
| | | Ivory Coast | Belgium | Rwanda |
| | | Hong Kong SAR | Botswana | Senegal |
| | | Israel | Chile | Seychelles |
| | | Italy | El Salvador | Singapore |
| | | Korea (Rep. of) | Greece | Slovenia |
| | | Paraguay | Iceland | South Africa |
| | | Portugal | Ireland | Taiwan-China |
| | | Uganda | Luxembourg | Tanzania |
| | | Venezuela | Mexico | United Arab Emirates |
| | | | Morocco | United Kingdom |
| | | | Netherlands | |

*Source*: International Telecommunication Union

**Figure 3.6: Top Mobile Economies** (2000, millions)

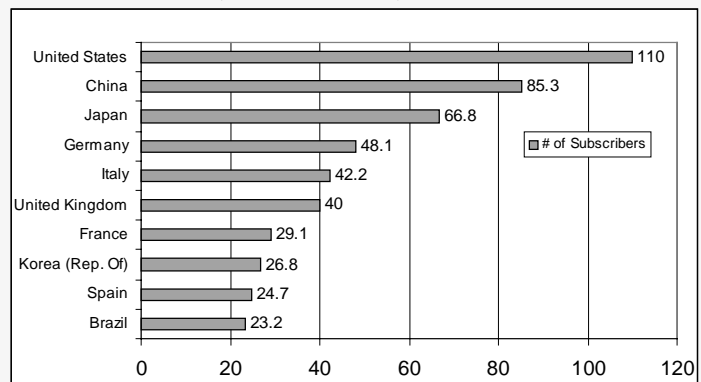

*Source: International Telecommunication Union*

While some expect the 2G peak to come sooner (between 2001 and 2002), others think 2G subscriptions are likely to peak in 2002/2003. The Yankee Group expects 2.5G subscriptions to start a slow dive around 2004, while Analysys and 3G Lab see such a decline from 2008 onwards (based on computer modelling of worldwide 2/2.5/3G markets).[70]  Most analysts seem to identify 3G's 'critical mass' period as likely between 2004 and 2006.  (See Figure 3.7)  Insofar as these estimates could be based on subscribership driven by factors like the availability of handsets, Figure 3.7 offers a compelling illustration of the potential dynamic between the various 2G, 2.5G, and 3G technologies.

**Figure 3.7:  Western European Cellular Users by Technology, 1997-2006[71]**

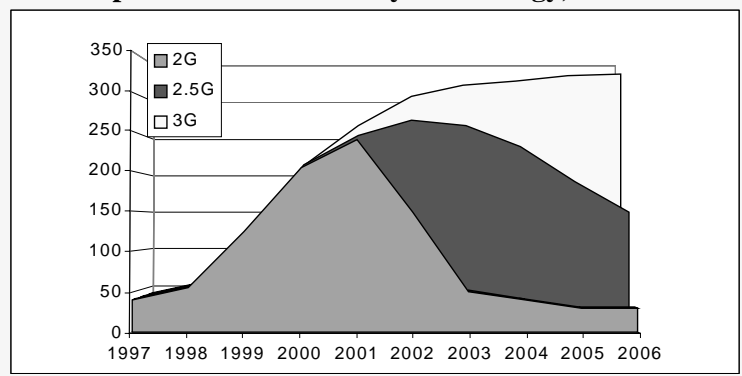

Source: *The Yankee Group, 2001*

Some believe the demand for mobile internet will come from corporate clients requiring mobile e-mail, Intranet, customer profiles, credit details and stock prices.  Others think the transactional capabilities of 3G will ensure the revenue-generating potential.  Total mobile internet subscribers are estimated to reach nearly 177 million by 2005.  (See Figure 3.8)  Corresponding mobile internet revenues are expected to grow from $5.3 million in 2000 to $3.8 billion in five years.[72]

**Figure 3.8:  Mobile By the Numbers: Penetration 2000 – 2005 (millions)**

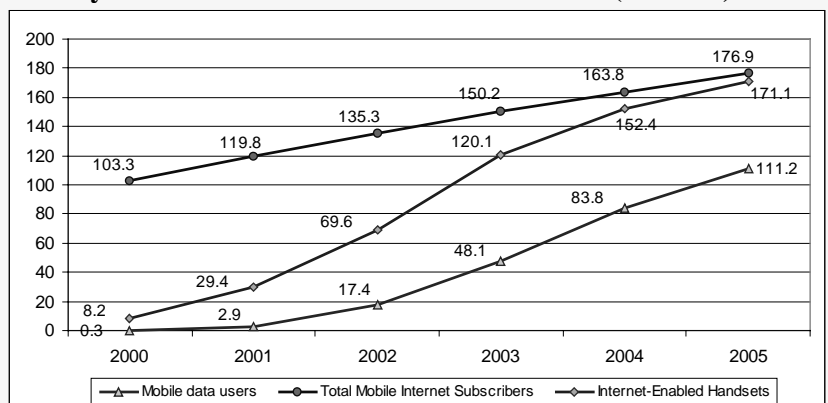

Source: *Forrester Research*

Mobile connections in Western Europe are expected to grow from 154.1 million connections in 1999 to 325.3 million connections in 2004, and penetration is forecasted to increase from 40% in 1999 to about 89% in 2004.  (See Table 3.4)  Prepaid customers, as mentioned above, are most likely to drive growth in the forecasted period.  Lower prices for voice service will also drive connection growth as more cost-conscious consumer customers are targeted.   A recent study into the service revenue opportunities over the next decade for the 3G mobile market conducted by the UMTS Forum predicts a Compound Annual Growth Rate (CAGR) for 3 key 3G services – namely Customized Infotainment, Mobile Intra/Extranet Access and

---

[70] *"Wireless: Riding its luck into 3G"*.  Mobile Matters, February 2001. p.53.

[71] *"3G in Europe:  Expensive but Essential"*.  The Yankee Group.  The Yankee Report Vol.5 No.8 - June 2001.

[72] McCarthy, Amanda. *"Mobile Internet Realities"*.  Forrester Research Report,  May 2000.

Multimedia Messaging Service – of over 100% during the forecast period, with total revenues for these 3 forecasted services of over US $164 billion by 2010.[73]

**Table 3.4:  Summary Forecast for Mobile Service in Western Europe (to 2004)**

|  | 1999 | 2000 | 2004 |
|---|---|---|---|
| Total Connections (thousands) | 154'112.3 | 211'862.0 | 325'283.0 |
| Analog Connections (thousands) | 5'704.7 | 3'696.9 | 36.7 |
| Digital Connections (thousands) | 148'407.5 | 208'165.1 | 325'246.3 |
| Prepaid Connections (thousands) | 75'294.9 | 117'899.1 | 198'590.1 |
| Postpaid Connections (thousands) | 78'817.4 | 93'962.9 | 126'692.9 |
| Total Service Revenue ($thousands) | $64,048,845.2 | $84,558,884.5 | $139,899,264.0 |
| Total Data Revenues ($thousands) | $2,150,111.5 | $6,142,510.1 | $45,608,645.9 |
| Total Average Revenue Per Unit | $521.4 | $462.1 | $439.4 |

*Source*: Gartner Dataquest (May 2000)

How the various data and penetration forecasts will ultimately translate to revenues, specifically through the facilitation of channels like mobile commerce, remains to be seen.  Based on the information provided below in Table 3.5, it appears that by 2005, Asia will be quick to adopt m-Commerce (generating revenues of US$9.4 million, nearly 60% of which will be led by Japan), followed closely by Western Europe (generating revenues of US$7.8 million), and more slowly trailed by North America (generating revenues of US$3.5 million, 94% of which will be US-led).  By 2005, it appears that US$22.2 million dollars of revenue will be generated globally as a result of transactions made possible by mobile devices.

**Table 3.5:  Global Mobile Commerce Revenues, 2000 - 2005 (USD millions)**

| Region | 2000 | 2001 | 2002 | 2003 | 2004 | 2005 |
|---|---|---|---|---|---|---|
| North America | 0.0 | 0.1 | 0.2 | 0.7 | 1.8 | 3.5 |
| Western Europe | 0.0 | 0.1 | 0.5 | 1.7 | 4.6 | 7.8 |
| Asia | 0.4 | 1.3 | 2.6 | 5.0 | 7.4 | 9.4 |
| Latin America | 0.0 | 0.0 | 0.0 | 0.1 | 0.2 | 0.5 |
| Other | 0.0 | 0.0 | 0.1 | 0.2 | 0.4 | 1.0 |
| Global | 0.4 | 1.5 | 3.4 | 7.6 | 14.5 | 22.2 |
| US | 0.0 | 0.1 | 0.2 | 0.6 | 1.7 | 3.3 |
| Japan | 0.4 | 1.2 | 2.1 | 3.5 | 4.5 | 5.5 |

*Source*: Jupiter Research[74]

## 3.6   3G Licensing Policies

The magnitude of 3G 'hype', aside from its focus on forthcoming services, is best exemplified by the debate over allocation methods for the scarce resource desired by every operator - spectrum.  W-CDMA networks will operate in a new range of frequencies higher than most 2G systems, and thus 3G mobile wireless networks have ushered in a momentous new round of spectrum licensing.  Therefore, any comparative view of 2$^{nd}$ generation GSM with 3$^{rd}$ generation IMT-2000 is incomplete without addressing this costly aspect of mobile roll-out:  license acquisitions.  Any operator with an established GSM network and a stake in mobile markets has been required to obtain a 3G spectrum license. With established GSM networks and shares of mobile markets, operators have had little choice but to join in the race.

---

[73] Eylert, Bernd Dr.  *"UMTS:  Making Mobile Multimedia Happen for Every Nation"*.  UMTS Forum, UK.  "Policy and Development Summit" ITU, EY-p.2.

[74] *"Global Market Statistics for Mobile Commerce, Part II"*, (accessed August 2001) Link: http://www.canvasdreams.com/viewarticle cfm?articleid=943.

In terms of license allocation methods, two have thus far been amongst the more prominent: auctions and beauty contests; the differences inherent in these processes have brought about tremendous variation in the prices associated with spectrum.

Auctions have been supported for the full transparency they bring to the allocation procedure, and for the weight they give to the 'dependable market' as selector of 'winners'. Price is seen in this context as an objective selection criteria, and one that is supported on the assumption that the money raised is actually close to the real economic value of such the license in question. The fact that price is directly oriented upon demand somehow lends credibility to even the most astronomical of valuations (at least according to some die-hard economists), giving the impression that risk is somehow mitigated amidst market trends justifying extremely promising uptake of mobile services in coming years. Finally, auctions are purported to offer more flexibility for the operators' roll out, coverage and market development. Some economists (clearly proponents, for example, of the aforementioned UK auction) believe that the enormous upfront costs of buying licenses have zero impact on the future prices 3G operators will be able to charge their customers; they are perceived to be the 'sunk costs' that operators should simply absorb as part of their strategy.

Beauty contests, on the other hand, are easier to follow, as well as more malleable in terms of being used as tools toward the implementation of special regulation (social or regional policy) goals. They grant more control for guidance to the regulator of the process vis-a-vis its' result, and also give a more flexible definition to the licensing object. According to an article last year in Red Herring Magazine, license allocation based on merit and not price tag is one that is unquestionably favored by industry. "… Industry favors this approach, fretting that huge license fees will slow deployment of the costly 3G infrastructure and hold back mass adoption as the added cost is passed on to subscribers."[75]

### 3.6.1    The European Experience

The race to 3G is undoubtedly about spectrum, and it is notable that this priority was not as controversial when GSM was being prepared for deployment. Much of this has to do with the fact that the majority of GSM licensing was executed by PTT's in a beauty-contest fashion. This past year alone, however, European network operators have forked over in excess of $100 billion for spectrum in the race to offer next-generation mobile services, with the hope that 3G will be a revolution toward strong growth and market stability. It became clear, particularly through the auction method of license allocation, that incumbents were by and large unprepared to give up their market positions in mobile telephony, at least in main European markets.

In the United Kingdom, five companies committed to paying a total of £22.5 billion ($35.4 billion). In Germany, six companies committed to paying DM98.8 billion ($45.85 billion). The exorbitant prices in the United Kingdom and Germany were determined in the end by how high prospective new entrants were prepared to bid. For smaller operators in smaller markets, the consolation was that the auctions left no funds available for smaller markets, allowing for smaller operators to be left alone at least in the short term. Even the large operators would hit limits of financing after committing huge amounts in the main markets.

On the surface, the auction model seems to be a great way for governments to hand out temporary monopolies on radio frequency, leaving the free market's 'invisible hand' to point to the 'right price'. However, the burden of responsibility for operators' incurred costs and the probabilities for operations in the 'red' has potentially dire consequences for the seamless integration of 3G service offerings around the world. Is it so unlikely, after all, that the high prices of licenses in some countries will not spill over on the countries that decided to part with their airwaves at more down-to-earth prices by adopting the 'merit-based' approach? A major feature of GSM, after all, was that it was possible to harmonize pan-European deployment in a way that did not compromise ultimate price offerings for the customers.

The bottom line however, as most see it, rests on the how much the costs incurred by operators will affect the average prices that must be charged to end-customers, such that operating expenses can be absorbed. (See Figure 3.9) Based on the graph below, it is interesting to note the 'wave' of the price trend over time from top to bottom, as countries are listed in the order that they allocated their licenses. While to some it may appear suspiciously similar to the volatility of recent telecom market sector conditions, to others it is rationalized as deliberate and proportional to the target market opportunities of the respective nations. The Economist, for one, seems to believe that regardless of the price peaks, mobile wireless services will not be inhibited "because the indebted winners would 'have the strongest possible incentive to roll out new services to recoup their money as fast they can.'"[76]

---

[75] Cukier, Kenneth and Hibbard, Justin. "*Spectrum Shortage*". <u>Red Herring Magazine</u>, September 1, 2000. (accessed October 2001)
    Link: http://www.redherring.com/story_redirect.asp?layout=story_generic&doc_id=RH1940013794&channel=70000007.

**Figure 3.9: Average Cost of 3G License Per Population**

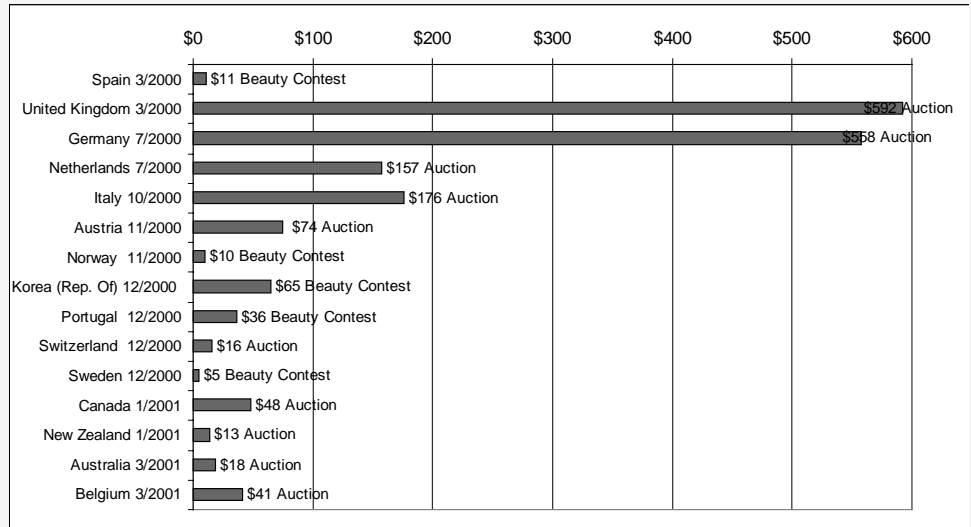

Source: *International Telecommunication Union*

From another perspective, the high prices paid for licenses reflect simply an intense and artificially-supported demand based on restricted supply, not taking into the account the profit potential of 3G spectrum after the costs of deploying necessary network infrastructure are met. Dresdner Kleinwort Benson declared of auctions: "a capital constraint has been created, inhibiting the growth prospects of the 'mobile multimedia society' and elevating the business risks."[77] Certainly, the goal of making ample amounts of spectrum available for industry as economically as possible is somewhat conflictual vis-à-vis governments' goals to maximize incoming revenues. The allocation of 3G licenses challenges governments to mediate between divergent public interest objectives: cashing in on their role as arbiters of radio spectrum, versus promoting competition and distributing the resource.

Is it feasible or fair to look at the financially ravaged operators as simply the bearers of some rather high 'sunk costs'? In other words, are their expenditures simply to be absorbed into normal operating costs? Is it realistic to consider them the unsuspecting victims of next-generation technologies, without assuming that consumers will be spared this tremendous cost burden, as depicted in Figure 3.7. Surely, operators will have to postpone widespread service offerings until scale economies are applicable to relevant equipment. How, without some price breaks amidst exorbitant roll-out costs, will 'winners' be able to "recoup their money as fast as they can"?

According to Martin Bouygues, CEO of Bouygues Telecom, operators face a choice between a fast death and a slow death: "… if they don't secure a license regardless of their price, the stock market decimates the company; if they win, the company bleeds itself over the license's lifetime (usually 15 to 20 years) as it struggles to make a profit."[78] Essentially, they are 'locked in' to making tremendous expenditures. Concerns over the ability of telcos to make reasonable ROI (return on investment) have resulted in a significant reduction in the availability of investment funds, which has in turn increased market angst. This reflects the makings of a vicious cycle, which is compounded by articles and commentary comparing 3G as a potential rival to the now-defunct Iridium mobile satellite systems project.[79] Surprisingly, however, 40% of respondents in a survey of operators conducted by the ARC Group believed that the 3G-licence auction process would have no effect on the rollout of next generation networks.[80]

---

[76] Cukier, Kenneth and Hibbard, Justin. "*Spectrum Shortage*". <u>Red Herring Magazine</u>, September 1, 2000. (accessed October 2001) Link: http://www.redherring.com/story_redirect.asp?layout=story_generic&doc_id=RH1940013794&channel=70000007.

[77] Ibid, Link: http://www.redherring.com/story_redirect.asp?layout=story_generic&doc_id=RH1940013794&channel=70000007.

[78] Ibid, Link: http://www.redherring.com/story_redirect.asp?layout=story_generic&doc_id=RH1940013794&channel=70000007.

Interestingly enough for free-market optimists, it is amidst those countries in which licenses have been awarded on merit (as opposed to market-based solutions) that services actually appear set to start sooner. And certainly, the comparatively smooth roll-out of GSM in the early 1990's confirms the hypothesis underlying this point. Such countries include Finland, Sweden, Japan, and Korea. Operators in these countries have the luxury of using the financial resources that they did not have to expend on the acquisition of licenses, for building out infrastructure for 3G services. (See Section 4.3.2 for further discussion of deployment costs).

It is ironic, given a climate characterized by mistrust of regulators and government intervention, that the example of Japan illustrates a country poised to provide services faster (and potentially cheaper) due to exactly those interventionist policies that run counter to 'free-market principles'. Japan's calculated bestowal of three licenses to transmit voice and data in unoccupied frequencies upon its three incumbent operators – J-Phone, KDDI, and NTT DoCoMo – was perhaps just what the country needed to give it a chance at a decisive lead over western counterparts. None of these operators had to pay up-front fees; they pay only radio-usage per subscriber per year, which add up to nothing much compared to the soaring auction prices in Europe. Naturally, operators in such an enviable position can defer capital that would otherwise go to the government, and invest in equipment, network-building and speedy service deployment. NTT DoCoMo's launch delays, though still less far off than others, are still, however, cause for concern for 3G roll-out in general. Concerns are valid, as earlier predictions about the range of 3G services were premature.

As time passes, the wide divergence in the results of license allocation methods becomes more and more prominent; most recently, the Liechtenstein-based Telecom FL actually decided *not* to exercise its right to a free UMTS license in the country, saying that it was unhappy with the terms of the license and thereby becoming the first company to decide not to accept its 3G license.[81] The case of Hong Kong is also an interesting one: a royalty-based payment scheme was recently introduced, intended to minimize the financial burden on operators by creating a schedule of minimum payments, which minimizes the government's credit risk, but still allows it to share in the potentially lucrative aspects of the 3G business. Under the scheme, each licensee in Hong Kong will pay the same percentage royalty on its future network turnover; this represents a compelling and thus far unique compromise. It is also remarkable to note that Finland actually issued their licenses free, representing a far cry from the Germany and United Kingdom auctions. It is as yet uncertain how the consequences of these diverging license allocations methods will be manifested in the IMT-2000 marketplace.

### 3.6.2    The American Experience

"We Americans are a jaunty and self-satisfied bunch, inclined to believe that if it ain't happening here, it ain't happening anywhere … yet there's one critical area of online technology where we're getting smoked: wireless."[82] – James Daly, Business 2.0.

"Industry executives and analysts [in the U.S.]… [are accusing]… the government of imperilling innovation, consumer choice and economic growth by failing to open the airwaves."[83]

It is of particular interest to elaborate briefly upon the American stance toward spectrum allocation, for the current implications it represents for global uniform frequency utilization and the American mobile wireless market are not insignificant. Although it is not untrue that American technological prowess rests on its ability to roll out the next generation of services, the federal government has certainly yet to deliver what the industry needs most to realize its future: the rights to transmit signals through its airwaves. Prior to 1993, federal regulators would accept applications from companies looking to use the public airwaves for things like television broadcasting or radio communications. If the proposed use was deemed to serve the public interest, and if there were no superior proposals from rival companies, the government simply granted the license. After 1993, when Congress decided that the airwaves could be better allocated through a more free-market-style auction process, licensing spectrum became a multi-billion-dollar government business.[84]

[79] "*The Winding Road to 3G*". eWeek, ZDNet Magazine. April 30, 2001. (accessed October 2001) Link: http://www.zdnet.com/eweek/stories/general/0,11011,2711432,00.html.

[80] "*Operators Express Concern Over Handsets*" Arc Group, January 16, 2001. (accessed August 2001) Link: http://www.arcgroup.com/ press2/cut_concernhandsets.htm.

[81] "*Company declines free 3G license*". CellularNews.com July 31, 2001. (accessed October 2001) Link: http://www.cellular-news.com/cgi-bin/database/archiveresults.cgi?week=161.

[82] Daly, James. "*Stateside Wireless Gaffes*", Business 2.0, October 9, 2000. (accessed July 2001) Link: http://www.business2.com/articles /web /0,1653,15070,FF.html.

[83] Goodman, Peter S. "*A Push for More Frequencies*", Washington Post, February 28, 2001. (accessed August 2001) Link: http://www.washtech.com/news /telecom/7918-1.html.

The FCC has thus far postponed three times an auction of airwaves initially planned for last October, to allow bidders to sort out the spectrum's value. This is a process fraught with overlapping claims from television broadcasters. Auctions now are not likely to occur before 2003. Current spectrum holders, including UHF television broadcasters and the Department of Defense, will continue to resist the re-allocation of these airwaves until they can find a way to monetize them.

Unlike Europe, the U.S. did not designate particular blocks of spectrum for 3G wireless networks; the auctions to come will sell off spectrum in the 700 MHz band, which owners will be able to utilize in a variety of ways. Despite President Clinton's executive order in the Fall of 2000 directing federal agencies to identify and make available new spectrum for the oncoming wave of sophisticated new services, the most attractive slices continue to be controlled by the Defense department. The radio bands in the U.S. thought to be most suitable for 3G are controlled in part by the Pentagon, which uses them for a variety of purposes, including communications with intelligence-gathering satellites. Coaxing current occupants of these slices of radio spectrum is an extremely weighty task; in April, the Department of Defense reported that it would take "as long as 2010 for non-space systems and beyond 2017 for legacy space systems to vacate the relevant spectrum… band-sharing is not an option [for security and interference reasons], nor is relocation unless the wireless industry make comparable spectrum available and foots the bill for moving costs, which could total $4.3 billion."[85]

Apparently, federal authorities have made available only about half as much spectrum as their French, British and Japanese counterparts. Not only does the U.S. have half the available spectrum of most other countries, but it also has a cumbersome spectrum cap of 45 MHz per market, per carrier. As a result, major American wireless carriers are now in the midst of a fierce lobbying campaign for new frequencies, while calling for an end to the federal limits on how much spectrum can be owned in a single market. "According to the CTIA, the number of minutes used by wireless customers multiplied by a factor of 13 from 1993 to 2000, while the amount of spectrum the government released for use less than tripled."[86] While some use this as point of departure against the spectrum cap argument, others see this logic as flawed, given the enhanced spectrum efficiencies that digital technologies help to create. If nothing else, this signals that maintaining artificially imposed caps on spectrum ownership may well have dire consequences for the strategic positioning and development of 3G-associated content, applications, and infrastructure providers.

Unlike in Europe, with its uniform mobile wireless standard based on GSM, North America (and the U.S. in particular) currently uses TDMA, FDMA, GSM and CDMA; with these four existing platforms, the path is nothing if complex. Essentially, North American wireless providers must gamble on which 3G platform will be most advantageous to implement. Both CDMA 2000 and WCDMA still have strong potential.[87]

---

[84] Glasner, Joanna. "When Air Isn't Free", Wired News, September 12, 2000. (accessed August 2001) Link: http://www.wired.com/news/print /0,1294,38669,00.html.

[85] "Investigation of the Feasibility of Accommodating the International Mobile Telecommunications (IMT) 2000 Within the 1755-1850 MHz Band", Defense Information Systems Agency, February 9, 2001. (accessed August 2001) Link: http://www.disa.mil/d3/depdirops/spectrum/Contents/imt-2000report/ ExecutiveSummary.pdf.

[86] Goodman, Peter S. "A Push for More Frequencies", Washington Post, February 28, 2001. (accessed September 2001) Link: http://www.washtech.com/news/telecom/7918-1.html.

### 3.6.3 The Asia-Pacific Experience

The Asian experience with spectrum allocation has been somewhat less problematic than the North American or European, mainly because there have been fewer concerns of overlapping 3G spectrum with existing allocated frequencies, and because spectrum has not been fetching the types of exorbitant sums as seen in Europe. Governments have learned from Europe's experiences, and have been modest in their proposals for spectrum license fees; in turn, operators have been cautious about network construction costs and time scales. Governments in Asia/Pacific have simply been less eager to maximize revenue from awarding for IMT-2000 licenses; as a result, they have been far less expensive.

New Zealand's auction of 2G and 3G spectrum, which started the ball rolling for the license process in Asia/Pacific, netted final bids totalling NZ$133.58 million ($59.6 million) in January 2001; it was the longest licensing auction at the time (it started in July 2000), and it led to some of the cheapest licenses allocated. Compared with the number of 2G users, the total of the proceedings represents US$35.5 per user, a far cry from the amount raised in some European countries.[88] Singapore's 3G licensing process recently concluded without any competitive bidding, with winners walking away with 3G licenses by paying S$100 million (US$55.6 million) each. In Australia, the total cost of 3G licenses was just 8% more than the A$1.08 billion (US$543 million) reserve price.

Accordingly, Gartner's research has revealed that 3G license cost in Asia-Pacific is about 10 times less expensive than that in Europe.[89] Japan represents the most attractive market for 3G development, as 3G licenses there have been issued at no cost even though cellular ARPU (average revenue per user) figures are among the highest in the world. Japan also currently leads the race to provide mobile Internet access in Asia/Pacific, with South Korea close behind. Of the region's mobile wireless Web service subscribers, 75% are in Japan and 23% in Korea. The remaining 2% of users are spread among all the other key markets in Asia/Pacific.[90]

Countries in Asia-Pacific are at different stages on the evolutionary path towards 3G. Broadly speaking, countries like Japan, South Korea, Hong Kong, Taiwan, Singapore, and Australia have been likely to be '3G early adopters', as they already enjoy high-cellular penetration rates and well-developed mobile wireless markets. On the other hand, less well-developed countries in which 2G demand has yet to be met – such as China, Thailand, the Philippines and Malaysia – are more likely to constitute a '2nd wave' of IMT-2000 adoption.

## 4    Comparing and Contrasting the Development of GSM and the Road to IMT-2000

The success of GSM in Europe was contingent upon a number of factors, not the least of which was the early and timely coordination of industrial actors, the creation of a full specification platform which allowed players to tailor their networks/services to different markets (without losing compatibility), the accessibility of essential technology, and strong political support vis-à-vis spectrum allocation, standardization efforts, and a regulatory environment conducive to competition. Other key factors included, of course, the expandability of the system (in evolution towards GPRS and EDGE), the self-organization of the mobile operators (into bodies like the GSM Association), and the creation of the open common platforms, which fostered competition not on systems, but on equipment and services. This helped to bring about the possibility of creating a mass market complete with low service tariffs and options for cheap equipment. Where GSM has succeeded, the groundwork for IMT-2000 has been laid; however where certain aspects of GSM's development cannot be compared with today's IMT-2000 'issues', it appears that the success of the 2nd generation Pan-European system is not to be taken for granted.

---

[87] Sprint PCS and Verizon Wireless are conducting field trials of CDMA 2000 1XRTT, in the hopes to be the first ones to roll out 3G in the US. VoiceStream and AT&T are taking the WCDMA route. From a standards migration perspective, the consequences of AT&T's decision last year to adopt GSM technology and its evolutionary pathway to WCDMA away from EDGE in the U.S. have been very significant; AT&T has had to build a second network (overlay) based on GSM technology and follow the GSM 3G pathway to 3G. (This is based on the assumption, however, that EDGE is considered to be a 3G solution equivalent in perception to W-CDMA.) AT&T and their TDMA partners will essentially be capping their investments in their TDMA networks and deploying new base stations at their existing cell sites. Qualcomm, which would appear to have the most to lose (on the surface) from this W-CDMA dominance, has certainly not been taking these challenges sitting down; Verizon and Sprint PCS are also staunch defenders of cdma2000. It should be noted however, that Qualcomm earns approximately 4% royalties on all types of CDMA products.

[88] Bidaud, Bertrand. "*First Asia/Pacific 3G Auction Completed: Gartner Dataquest Analysis*", Telecommunications Televiews, Issue 4, January 25, 2001.

[89] "*Asia-Pacific's low 3G licensing costs benefit 3G development*", CMPnetAsia Team, AsiaTele.com, April 19, 2001. (accessed September 2001) Link: http://www.asiatele.com/ViewArt.cfm?Artid=8650&catid=6&subcat=62.

[90] Johnson, Geoff. "*Lessons in Mobility from Asia/Pacific*". Gartner Group Research, July 12, 2001.

Now that over twenty countries have awarded 3G licenses (56 across Western Europe, to be specific) and over 70 3G infrastructure contracts[91] have been signed, it is reasonable to believe that Europe is well on its way to offering 3G services. Concerns are spread over a very broad spectrum of doubt – encompassing fears that 3G will not be the financial success it was promised to be, and even perhaps that 3G may not make it to the market. Analysts from the Yankee Group believe that 3G will undoubtedly come to market, and that from a service perspective, do so even arrive this year. Although anticipated infrastructure and handset delays are expected, coupled with rather leisurely emergent returns on investment for operators, these factors will at best postpone the adoption of 3G rather than signal its end. Below are some factors which characterized the development of GSM, and which are very likely to be relevant for that of IMT-2000 (and hence UMTS).

## 4.1 Lessons from GSM that Apply to 3G

### 4.1.1 The Shifting Dynamic of Major Players

As mentioned above, prior to the liberalization in the 1990s, European telecom markets were firmly controlled by national governments and their respective PTT monopolists.[92] Although the European Community in 1993 agreed to fully open EC markets for telephone services by the start of 1998, most national governments opted to extend their state monopolies until the 1998 deadline to get themselves fit for competition.[93] One could argue that this period contributed significantly to the focused development and deployment of the GSM system in Europe.

Illustrating the general complexity of the case of GSM, it is necessary to bear in mind the fact that the actors involved in the GSM deployment process changed considerably over time. While international deliberations began on the level of the PTT representatives, the final bargain was struck by national governments. Supranational institutions and private corporations had played key roles even before the general agreement was reached, but their importance grew substantially once it came to implementing the framework, determining technical specifications and rolling-out service.

The process of defining UMTS – as ultimately a component under the IMT-2000 umbrella - is very much influenced by the fact that it is a service and a system emerging in the aftermath of GSM's success in Europe. The emergence of the IMT-2000 vision as a global one has been facilitated precisely because the GSM vision was pan-European (and successful). Although both concepts were born not very far apart in the 1980's, one had to develop before the other could be realized. And certainly, the European Commission had its GSM interests to protect while the ITU was planning IMT-2000, and whilst the future for GSM in the 3G context was still under determination. Accordingly, the semblance of a framework for a range of relevant partnerhips, consortia and interest groups gradually emerged by the time the 'negotiations' for technical specifications of the 3G standard started to take place. The ETSI, specifically, was instrumental in asserting the continuing crucial position of the GSM system as well as itself, as the transition to UMTS-oriented goals was taken underway.

"Given that UMTS is in many respects a continuation of the GSM process with corporate actors having assumed some of the roles previously played by the public sector, the question whether Europe's success in mobile wireless technology resulted from a particularly favorable industrial and political constellation, or whether it is the result of a robust and replicable ICT standardization process, still remains."[94]   The European Commission's previous concept for GSM as the result of a mutually beneficial cooperation between the public sector and private-sector consortia (in creating technical standards) appears to be working again for UMTS.   The same coalition of equipment manufacturers, network operators, telecommunications administrations and supranational institutions that paved the way for GSM has lent its support to UMTS. "Contrary to GSM, however, efforts are led primarily by manufacturers and private operators with supranational institutions focusing on the provision of fora for cooperative exchange and... legal backing."[95] It is at least in part due to the previously established unity and strength of old GSM European manufacturers and operators (as well as institutional structures put in place by the EC) that UMTS has had the clout that it had in broader IMT-2000 standards negotiations of the early 1990's.

---

[91] Roberts, Simone. "*3G in Europe: Expensive but Essential*"; Wireless/Mobile Europe, <u>The Yankee Group</u>, Report Vol. 5, No. 8 – June 2001, p.2.

[92] This is not to say that in the early 1990's, European business leaders were not aware of the potential benefits of liberalization. The European Community in 1993 agreed to open fully the EC's markets for telephone services by the start of 1998, and at that time it appeared likely that only six of the EC's 12 member-states (Britain, France, Germany, Italy, Denmark and the Netherlands) would definitely meet the date. "Of …500 senior decision-makers in eight European countries [in 1993], 85% of the respondents thought that telecoms liberalization would reduce business costs, 91% believed that liberalization would improve the range of services available and 92% thought that competition would stimulate improved quality of service." From an article by Woolnough, Roger. "*Liberal does of Telecoms*", <u>Electronic Engineering Times</u>, CMP Electronics File. October 4, 1993, p.24.

[93] Beardsley, Scott & Patsalos-Fox, Michael. "*Getting telecoms privatization right*". <u>The McKinsey Quarterly</u>, January 1, 1995. p.3.

The absence of a political force (like the European Commission was for GSM) has been noticeable in this phase of determining the definitions of '3G', and there has been less pressure for standards harmonization as a result.  Operators from around the world with massive international operations have been fighting this time for 'global footprints', as opposed to 'pan-European' ones, and the upshot effect for supra-national institutions like the ITU has subsequently been a manifest accommodation for respective operators' 3G 'definitions'.   Uniting cellular standards for seamless integration of GSM in Europe fell far more conveniently under the jurisdiction of players (both governmental and private sector) who had good clout and did not hesitate to use it; applying the same pressures on a project of such global magnitude has been less feasible.

### 4.1.2    The Critical Role of Equipment Manufacturing

The manufacturers of handsets for cellular terminals have played a critical role – both in terms of delaying launch of new services and raising costs of roll-outs for operators.  This was true for GSM, and is likely to be true again for all IMT-2000 systems.  In 1992, for example, GSM terminals were still not available in commercial quantities, and their lack was a major reason for delays in the startup of commercial GSM services in Europe.  "Delays were costly for the industry, network operators and service providers alike... [for instance], German service providers were losing between between $4.5 million and $6.4 million worth of business each month..."[96]  Manufacturers of course have historically had an extremely high stake in the success of their handsets; in 1992, "around 90% of the total investment already made in GSM--estimated at around $1.2 billion- had come from the manufacturing industry."[97]

"In the past, the success of the handset has greatly contributed to the success of a mobile offering..."[98]  Part of the evolution of the European markets has been discernible in operators' use of handset subsidies.  Over the past decade, operators in Sweden, Norway, and Denmark have used handset subsidies to offset the high cost of GSM handsets, which in turn have likely translated into a higher subscriber acquisition cost model.[99] (See Section 4.1.3)  The importance of the role of handsets in the deployment of GSM (both in terms of functionality and cost) is compounded now in the 3G scenario.  While handsets for GSM are at this point highly regulated and certified[100] after years of 'touch and go' (problems of over-heating, problems with 'dual-mode', etc.), serious concerns for 3G handsets abound.

Many cellular operators believe that a handset shortage will in fact delay the launch of third generation mobile services.  There are serious worries that they will be delivered late and will perform worse than the GSM phones they are meant to replace.  According to a recent survey of operators by ARC Group almost 90% ranked non-availability of 3G handsets as the primary barrier to the successful introduction of next generation services. Many companies are still smarting from problems with the supply of WAP and GPRS handsets and fear that similar problems will affect 3G services with potentially disastrous consequences. [101]

---

[94] Bach,  p.18.

[95] Ibid, p.17.

[96] Williamson, John.  "*GSM bids for global recognition in a crowded cellular world*".  Telephony (Intertec Publishing Corporation). April 6, 1992. p.36.

[97] Ibid,  p.36.

[98]   "*3G reprieve for Japan handset makers*",  CNN.com,  April 25,  2001.  (accessed  August  2001)  Link: http://www.cnn.com/2001/BUSINESS /asia/04/25/tokyo.handsetreprieve/.

[99] "*Wireless/Mobile Communications Europe*",  The Yankee Report, Vol.2 No.4 - March 1998.

[100]  In 2000, the GSM Certification Forum (GCF) was launched, representing a completely independent programme that aims to implement, verify and monitor an entirely new global voluntary certification process for testing GSM handsets and terminals. Over 50 operators, representing a combined subscriber base of over 100 million customers, have signed Declarations of Participation in the GSM Certification Forum. In addition, all 11 of the primary GSM terminal manufacturers - providing more than 95% of all handsets/terminals sold world-wide - have signed Declarations of Participation.  "*Global Partnership to Benefit all - The Launch of the GSM Cerfitication Forum*".  GSM Association.  (accessed  August  2001)  Link: http://www.gsmworld.com/news/press_releases_44.html.

Critics says a complex development such as IMT-2000 requires a great deal more time to be completed and tested than the Europeans have allowed. 3G base-stations and telephone handsets have had to be created from scratch because of Europe's insistence on following its own version of the CDMA technology. Third-generation handsets will need to roam between 2G, 3G, GPRS and GSM networks in Europe, between PDC and wideband CDMA (W-CDMA) in Japan and between time division multiple access (TDMA)/code division multiple access (CDMA) in the Americas. There may also be a need for roaming between different implementations of the 3G standard, such as Wideband CDMA (W-CDMA) and CDMA-2000.

**Figure 4.1:  Western European Handset Shipment Volumes by Technology[102]**

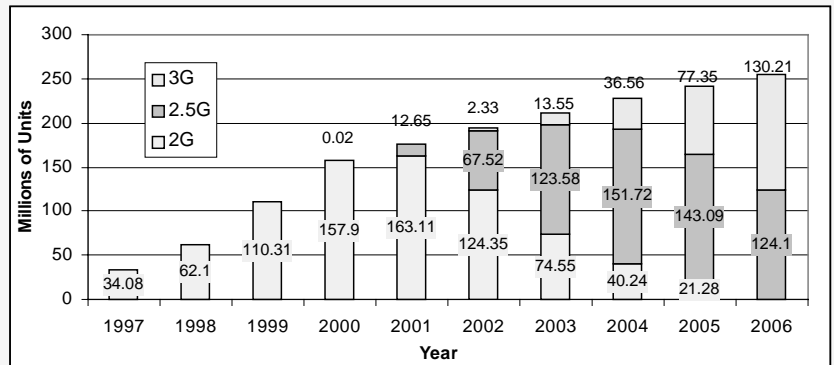

Source: *The Yankee Group, 2001*

As penetration increases, the potential number of new customers (handset sales) declines, with handset replacements constituting the majority of new sales. The Yankee Group forecasts that 74% of all handsets sold in 2001 will be replacement handsets. However, by 2006 mobile penetration is forecasted to be 85%, and accordingly 99% of all handsets sold will be replacements.

A good example of the echoing importance of handsets appeared in July 2001. Japan's NTT DoCoMo issued "an advisory" to owners of 100,000 Web-enabled P503i i-mode phones after finding they were unable to receive voice calls and email at certain geographical locations. DoCoMo also temporarily halted sales of Panasonic phones (made by Matsushita), while it identified which handsets were subject to the glitch by checking serial numbers.[103] A hold-up of this nature, or, for example, of GPRS handsets, affects replacement cycles as mobile users hold out for the new technology before replacing their handsets. The number of GPRS customers operators can hope to win by year-end will also be impacted.[104] "Similar delays are foreseen for WCDMA handsets. By the end of 2002, 2.33 million WCDMA handsets are expected to be shipped, constituting 1.2% of total handsets shipped. In 2005, the Yankee Group expects 3G and GPRS handsets to constitute 91% of all handsets shipped. (See Figure 4.1) The Yankee Group also believes that shipments of GSM-only handsets to western Europe will have ceased by 2006, since vendors will be keen to cease production of GSM-only handsets as quickly as possible to reduce production line costs and force subscriber migration to higher-generation products."[105]

---

[101] "*Operators Express Concern Over Handsets*" Arc Group, January 16, 2001. (accessed August 2001) Link: http://www.arcgroup.com/ press2/cut_concernhandsets.htm.

[102] "*3G in Europe: Expensive but Essential*". The Yankee Report, Vol.5 No.8 - June 2001.

[103] "*DoCoMo takes 2 million minutes to fix flawed i-mode phones*". Mobile Media Japan, July 11, 2001. (accessed August 2001) Link: http://www.mobilemediajapan.com/2001/07/11.html.

[104] With Nokia, the world's number-one handset vendor, announcing that its GPRS handsets will not be available until the third quarter of 2001, operators will be unable to secure a large enough number of handsets to effectively promote GPRS to the mass market before this date. Although Nokia remains optimistic about launching dual mode WCDMA/GSM handsets in the third quarter of 2002, other handset vendors have stated that handsets will not be available in large quantities until the second half of 2003. "*3G in Europe: Expensive but Essential*". The Yankee Report, Vol.5 No.8 - June 2001.

[105] Ibid.

### 4.1.3    Learning from the Numbers

*4.1.3.1    The Cost of Acquiring New Subscribers*

It has been said that GSM success is best observed in the context of escalating penetration rates and high subscriber growth.[106] High subscriber levels for GSM, however, did not necessarily equate to high profit margins: operators have to face the issue of higher subscriber acquisition costs (SACs) when attempting to attract the less profitable customers of the mass market.  In the context of examining the potential cost burdens carried by users of 3G mobile technologies, it is crucial to briefly consider the subscriber acquisition costs (SACs) associated with these consumers.  As services become increasingly broad-ranging both in terms of breadth and geographical reach, it becomes apparent that not all consumers (or rather, subscribers) are 'created equal'; in other words, different users become valued differently as a result of increasing acquisition costs.  This is inevitable, since as markets get more competitive, a general struggle around price offerings becomes discernible; discriminating, price-conscious consumers in a fickle market constantly impose pressure for better value for their money.   Therefore, new users are attracted to cellular at the expense of growing acquisition costs while yielding lower than average revenues in all but the longest of terms.

This is important to bear in mind precisely because predictions of the take-up of 3G are often postulated on the basis of 2G (GSM) penetration levels, ignoring the varying 'value' of the individual subscribers.  That subscriber numbers after a certain threshold (believed to be the first 10-20% of subscribers[107]) have an inverse relationship with revenues is a sad discovery since the days of 2G deployment; this is unlikely to change for the next generation, particularly as the market nears saturation.  "… in the first three to five years of a GSM network's life, given a buoyant economy, operators attract the more lucrative customers, but '…in their endless search for extra growth on top of that, they have to lower their expectations'."[108]  Many of the current users of data services tend to be 'early adopters'[109] and therefore not reflective of the typical profile for a subscriber; in turn, revenue growth prospects are reduced as penetration increases.  This yields somewhat of a counter-intuitive result, to what would otherwise appear as an optimistic growth scenario.

2G revenues are also expected to help fund the development of 3G networks and services, such that any decline in per-subscriber revenue hits not only current profit expectations, but also future investment planning.  This is one more reason why the study of GSM deployment and penetration is important for understanding 3G.  It is essential to understand such links between generations, even though upcoming service offerings may be completely new and unlike what has been offered before.  Looking at the positive aspects of 2G GSM penetration and growth cycles does not necessarily mean that comparable absolute revenue growth should be expected from 3G.  There is no certainty as to how people will react to data service availability, regardless of how optimistic cellular penetration forecasts appear to be, nor is their certainty as to the 3G-specific threshold above which IMT-2000 SACs will escalate.

Although penetration rates in Western Europe have increased greatly, the subscriber acquisition costs incurred through subsidizing less profitable customers seem set to remain high as the market approaches saturation point.  Subscriber growth is also expected to slow, which analysts consider to be the potential result of compounding competitive pressure as operators fight more aggressively for new subscribers.  With the prospect of this competition, operators will inevitably face an increase in subscriber acquisition costs as they attempt to woo subscribers from their competitors.  This was the experience of Western European operators in the first half of the 1990s.

*4.1.3.2    Subscriber Revenues*

The principal driver behind the development of mobile technologies is the potential value that will be created by mobile data services, especially via the mobile internet and m-commerce.  An important metric used to illustrate the effects of market penetration and saturation is 'Average Revenues Per User' (ARPU), which has been used extensively in assessing GSM market activity and forecasts.  Gartner Dataquest expects that ARPU in the context of gradual IMT-2000 deployment will begin to increase in Western Europe in 2003, at which point 'minutes of usage' will have increased sufficiently to offset a high influx of (expensive) consumer customers and a decrease in voice call prices. Operators will use the increase in non-voice traffic to counter lowering voice tariffs and to increase ARPU. For now, current data traffic accounts for only a small percentage of total revenues, on average about 7% in Europe. (See Figure 3.2) Data revenue is expected to grow, as voice ARPU has been declining steadily and is expected to continue to do so. To counteract this downward trend, operators are hoping to increase revenues through a dramatic increase in the usage of higher value data services. Some operators expect non-voice revenues to overtake voice revenues by 2004.[110]

---

[106] *"Wireless: riding its luck into 3G"*.  <u>Mobile Matters</u>, February 2001, p. 48.

[107] Ibid,  p. 49.

[108] Ibid,  p. 49.

[109] Early adopters with high average expenditures on technology will be the primary target market for new entrant 3G operators, whereas GSM-provider incumbents will likely capitalize on their existing customer bases, having probably already locked in many customers with their GPRS services.

As operators try to attract new subscribers (outside the initial 10-20%[111]), they may find that their ARPU indices may actually decline. Increasing the subscriber base has proven itself to be a double-edged sword in terms of a strategic move for operators, as pushing subscriber levels past this said threshold (in the GSM scenario) has helped lead to increased SACs. Forrester predicts that despite expected increases in mobile internet usage, ARPU for European mobile users will fall by 15% between 2000 and 2005, from 490 euros (about $448) to 419 euros (about $383).[112] That this ARPU forecast will indeed be relevant to IMT-2000 deployments is not certain, but it is surely an important point to bear in mind. The most crippling costs affecting eventual ARPU in most markets are likely to be those associated with handset subsidies, although advertising costs can be equally paralysing to operators' profit and loss statements.

### 4.1.4    Timeline for Deployment

**Figure 4.2:  GSM Timeline - 1982 to Present**

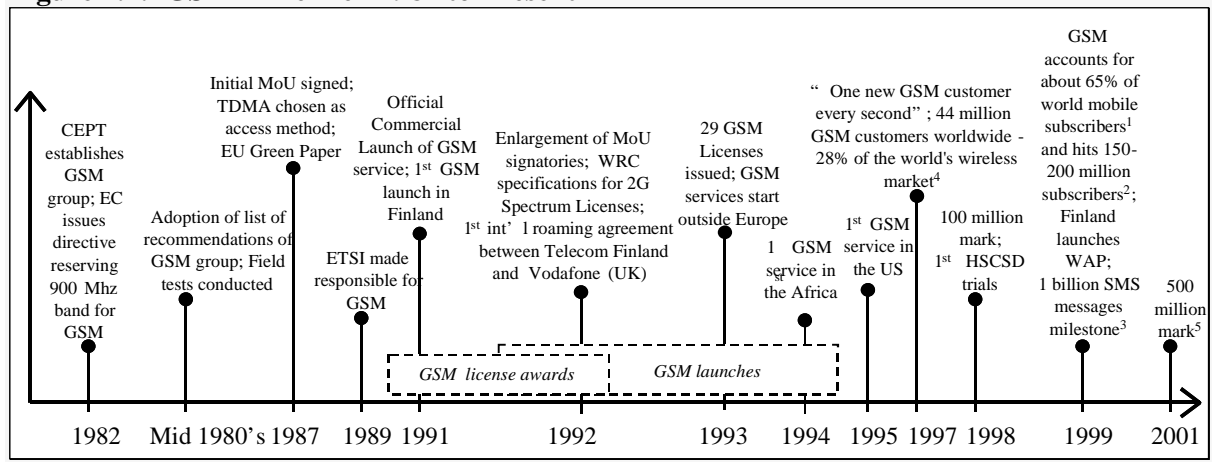

Note: [1]**See Link**:  http://www.gsmworld.com/news/media_18.html; [2]**See** "GSM Subscribers Hit 150 Million Mark" **Link**:  http://www.gsacom.com /news/gsa_020.htm **and Link**: http://www.gsacom.com/news/gsa_032.htm; [3]**See Link**: http://www.gsmworld.com/technology/sms_success. html;  [4]**See Link**: http://www.gsmworld.com/news/press_archives_14.html. [5]**See Link**:  http://cellular.co.za/gsmhistory.htm.
*Source:* International Telecommunication Union

A glance at the creation and evolution of GSM (See Figure 4.2) shows us that this was a system that took years to develop. Given its more recent success, the difficulties of GSM deployment of the early 1990's vis-à-vis troubles with equipment and legacy systems are often conveniently forgotten. The complex interplay between manufacturers of network and system equipment, the goals of governmental directives, operators' financial priorities, special interest groups, the demands of consumers, and the ultimate performance of service offerings – all brought together under the auspices of standard-setting organizations like the ITU – makes for a process which has turned out to be both time-consuming and extremely intricate. If anything, a healthy perception of the time frame necessary for deployment of any kind of cellular service is vital – not only for managing 'market' expectations, but for the purpose of managing expectations among consumers as well. WAP taught a valuable lesson to mobile internet enthusiasts about the virtues of patience; without it, the risk of dooming a technology to a bad reputation (that can only possibly be undone with great amounts of marketing expenditure) is increased.

---

[110] *"European Overview".*  Frost & Sullivan Research.  2001,  p.2-1.

[111] Percentages above this threshold imply a likelihood that subscribers be lower volume users. *"Wireless: Riding its luck into 3G".*  Mobile Matters, February 2001,  p. 49.

[112] Godell, Lars. *"Europe's UMTS Meltdown".*  Forrester Research Report, December 2000,  p.8.

In the same vein, a realistic perspective on the deployment of $3^{rd}$ generation systems is crucial. Many forget that 3G has been quite long in the making as well, although perhaps shorter in 'conceptual' timeline than GSM. (See Figure 4.3) Although perhaps the idea for wireless data delivery was conceived in general conjunction with that of wireless voice service, the actual processes for the creation of 2G and 3G respectively were carried out in mutually exclusive settings – at least until the ITU stepped in to create IMT-2000 as a standard designed to integrate and incorporate legacy 2G systems. Indeed, $2^{nd}$ generation networks (and GSM in particular) had to be deployed first before 3G could be realized, and today's perceived 'race' toward 3G reflects the harbored illusion of those who may not recognize the history of its creation. The 'race' is also arguably the creation of those preoccupied with the panic of recent spectrum prices. The next generation *is* on its way, but the time necessary for its smooth deployment is something that is not sufficiently accounted for in market analysis and the press.

### Figure 4.3: 3G Timeline: From 1989 to Present

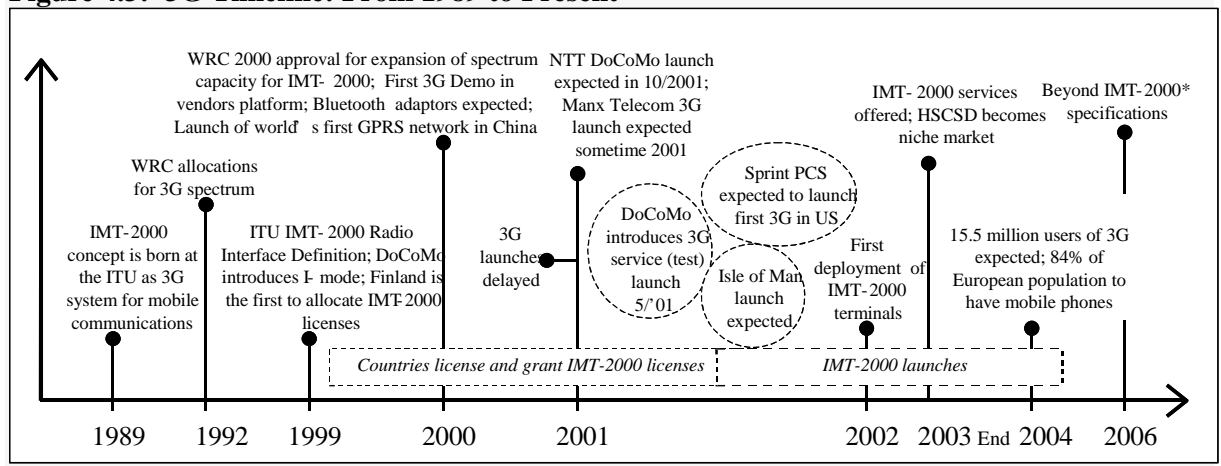

Note: [1] Further Forecasts for 2004: Number of mobile connections: 328.0 million; Users of SMS services: 203.6 million; Users of GPRS: 136.3 million; Users of circuit-switched data: 82.0 million; Users of 3G/UMTS: 15.5 million Users of EDGE services: 3.2 million. From "The Next Generation of Mobile Networks Poses a $100 Billion Challenge for Europe". Gartner Group Research. September 19, 2000.
\* "Beyond IMT-2000" is the terminology used to refer to what is otherwise known as "4G networks" by the ITU, as the transition from 3G → 4G is not considered to be a paradigm shift equivalent to that of 2G → 3G; Anything "Beyond IMT-2000" is most likely to be a continuation of 3G packet-based networks.
*Source*: International Telecommunication Union

## 4.2    Lessons from GSM that Don't Apply to 3G

### 4.2.1    A Harmonized Approach to License Allocation

When GSM was being developed, national governments were free to choose to whom a license would be issued - and with the exception of the UK - issued the first of their GSM licenses to their national PTT's. One could argue in this case that the success of GSM – particularly in its harmonized approach to license allocation – has not been replicable (or even applicable) to the global 3G case.

In Europe, in preparation for IMT-2000, each country regulator was given the responsibility of setting its own licensing conditions and procedures – and this has led to wide cost variations in the price of 3G licenses across Europe. The Council of Ministers and the European Parliament of the EU adopted a 'UMTS Decision' in 1998, aimed to ensure the availability of at least one inter-operable service in the EU, while leaving the characteristics of that service to concerned operators and suppliers. The corresponding 3G licensing conditions set by most country regulators ruled that all license winners should build their own networks, and specified a date by which network rollout must be complete and services launched.[113] This led, as we have seen, to massive volatility in the valuation of spectrum, as well as considerable doubt as to the financial viability of countless European operators. Given this, how responsible should the EC itself be held for the 3G 'difficulties' currently engulfing Europe?

---

[113] Fines for delayed network launch dates in some cases have been waived by regulators, one such example being Spain (due to launch by August 2001, now set for Q2 2002).

It is interesting to see that the US is only doing things slightly differently for 3G than they did during the development of GSM. There are differing perspectives on the American treatment of questions around 3G spectrum. "The lesson from GSM [the predominant technology for mobiles in Europe and much of Asia] is that we did it our way and we got left out of global roaming," said Leslie Taylor, president of a wireless consultancy in Washington, D.C.[114] The U.S., from the start, has been opposed to any measures that restrict competition or limit the flexibility of service providers to meet market needs, particularly in order to protect consumers from increased costs. The U.S. government has maintained thus far a rather cautious approach to 3G services in general, which opts for leaving the major details of wireless spectrum usage 'to the industry'.

Some clarification, however, was recently given to the matter of spectrum in the U.S. upon the announcement that President Clinton had issued an executive memorandum: "…urging federal agencies - including the departments of commerce and defense -- to work together to identify spectrum that can be used to implement 3G networks".[115] Therefore, the previously murky migration path to 3G mobile services in the U.S. is somewhat clarified, although it continues to fall well outside the boundaries of GSM migration and W-CDMA solutions. Negative perspectives on the questions of the U.S. deployment of 3G continue to abound: "…While the American laissez-faire approach to standardization and the resulting multitude of analog and digital standards created a competitive environment that put pressure on prices early on (leading to a rate of diffusion initially higher than Europe's), the American wireless market [following the logic of Metcalfe's Law]… is inherently limited in its application potential as a result of incompatibility of networks and market fragmentation."[116] However, others argue that the United States may in fact quickly find itself on a better path to 3G than anyone else: they have invested in a unique standard that can attain 'cdma2000 status' without the need for new spectrum. Depending on one's perspective, this could reflect the basis for a significant potential American 'recovery' in the race to 3G – and even cast previously-respected EC Directives for cooperation on technology standards in a darker light. Either way, however, the U.S. position continues to be exclusive and American operators will continue to face the challenges of global roaming plans.

### 4.2.2    The Underlying Philosophy of the Marketplace

Much, of course, is related to the underlying common philosophy of the marketplace at a given point in time. "Liberalization has not only led to the virtual disappearance of the requirement to use official European standards in telecom procurement, it has also dramatically increased the number of corporate players in the industry."[117] When GSM was coming around, the fear of interventionism on the part of government was not as acute, and gripes associated with, for example -'beauty contest' license allocation methods, were not necessarily perceived of as antithetical to creating 'free market' efficiencies. This, as shown below, is a big preoccupation in particular for American regulators. Although national governments in Europe in the 1990's were in the process of liberalizing and deregulating, the need for centralized, organized efforts on the part of the European Commission in Brussels to drive GSM were appreciated and encouraged. This is certainly not the case in present day.

### 4.2.3    Intellectual Property Rights (IPRs) and Limitations on Manufacturers

One crucial area in which the GSM experience may not be transposed on the development of IMT-2000 is that of the application of IPRs to the manufacturing of mobile equipment (i.e., handsets). This, contrary to many of the above points, reflects an area in which GSM failed, and in which it is hoped IMT-2000 will succeed. IPR policies, throughout the development and deployment of GSM, put severe limitations on the number of companies that were accredited with the right to manufacture GSM equipment. For GSM, there were about 20 companies that owned the essential technology necessary to realize GSM system.[118] This created not only a cap on the availability of handsets, but potentially aggravated the pace of successful deployments, creating imbalances in the industry vis-à-vis those who could profit from the system, and those who could not. Operators in countries who had all the makings for cost-effective successful deployments, including the skills and manufacturing base to produce their own equipment (i.e., Brazil), were forced to buy from the privileged few European manufacturers, and thereby impose their incurred cost burdens ultimately on their own customers. Certainly, this put a strain on the attainment of market efficiency for operators in non-accredited countries.

---

[114] "*3G Spectrum Allocation: The U.S. Leaves the Industry to Divide*." <u>WirelessReporter.com</u>, (accessed August 2001) Link: http://www.pervasiveweekly.com/issues/ pvw06082000.html.

[115] "*Clinton: Find Me 3G Bandwidth*". <u>Wired News</u>, October 13, 2000. (accessed August 2001) Link: http://www.wired.com/news/technology/ 0,1282,39451,00.html.

[116] Bach, p.17.

[117] Ibid, p.18.

[118] "*Modern Technology Transfer Approach*". <u>The 3G Patent Platform</u>, (accessed August 2001) Link: http://www.3gpatents.com/.

Industry is now seeing a paradigm shift in the complexities of intellectual property vis-à-vis the equipment manufacturing sector. It is believed that over a 100 companies/organisations will now own the technology (patents) necessary to realise a 3G system[119], reflecting a vast improvement over the past.

Similar IPR concerns can also be more broadly applied to manufacturers supporting competing standards – for example, W-CDMA and cdma2000. Protectionist, non-collaborative inclinations of manufacturers vis-à-vis the sharing of IPR's can result in the development of non-interoperable equipment, which in turn can lead to negative reverberations in a market for years to come. To illustrate this point, Motorola represents a good example in the GSM context because of its competing handsets with Nokia; it was consequently considered as something of a 'black sheep' in the GSM era. Indeed, the monopoly of manufacturers of GSM equipment was extremely difficult to penetrate by companies like Motorola. Although standardization issues in the 3G context are more global, and despite the fact that many numbers of corporations are now involved, two competing specifications groups still remain, and both are moving in their own respective directions although 'bridges' have been built between the two projects. One is the European-backed 3GPP[120], and the other is the (US-based) Qualcomm-backed 3GPP2[121]. Thus, despite the existence of hundreds of manufacturers, there is unlikely to be a solution to the broader IPR dilemmas until the efforts of 3GPP and 3GPP2 are effectively merged to make real collaboration possible.

This reflects more on the changing nature of the relationship between operators and manufacturers than on the provision of more widely available access to technology patents. Whereas this dynamic between operators and manufacturers in the GSM context could be characterized as more 'balanced' in terms of sector market 'drivers', the IMT-2000 context reveals a situation now in which operators' goals have since taken a back-seat to the goals of manufacturers. The strong position of operators in the early-mid 1990's, reinforced by their influence in ETSI, created for an environment in which operator concerns remained as primary drivers of industry activity. This was not contradicted by national manufacturers at the time, for they in turn were focused primarily on their traditional 'primary' clients – the national PTTs. Today, by and large, things have changed significantly (with the major exception of NTT DoCoMo, which has been a leader of the 'operator-led' paradigm). For the most part, manufacturers have proven themselves to be rather less motivated by issues of standardization- and more concerned with their own 'bottom lines'; the fact that discussions today about issues like interoperability are undertaken by 3G manufacturers is something of an indirect confirmation of their enhanced influence and position in the 3G value chain. This, if nothing else, signals the importance of renewing and strengthening not only increased participation of operators in 3G standardization, but also of greater collaboration between operators, manufacturers and content providers in an international forum.[122]

---

[119] "*Modern Technology Transfer Approach*". The 3G Patent Platform, (accessed August 2001) Link: http://www.3gpatents.com/.

[120] The world's leading telecommunications companies have come together and completed the definition of the 3G Patent Platform for handling the intellectual property rights associated with the 3G standards adopted in the ITU's IMT-2000 framework. Those concerned with GSM in the mid 1990's are fully aware of all the problems and difficulties associated with the licensing of GSM technology. Many companies could not enter the GSM market due to excessive GSM royalty rates, and because of all the obstacles in the complex minefield of negotiations with so many companies claiming ownership of essential patents. The 3G Patent Platform is an innovative technology transfer mechanism which introduces a quantitative approach as to what is "fair, reasonable and non-discriminatory" licensing conditions for essential patents. The 3G Patent Platform is about making the 3G technology more affordable to all players." The 3G Patent Platform, (accessed August 2001) Link: http://www.3gpatents.com/.

[121] The Third Generation Partnership Project 2 (3GPP2) is a collaborative third generation (3G) telecommunications standards-setting project comprising North American and Asian interests developing global specifications for network evolution to 3G. "3GPP2," which, like its sister project 3GPP, embodies the benefits of a collaborative effort (timely delivery of output, speedy working methods), while at the same time benefiting from recognition as a specifications-developing body, providing easier access of the outputs into the ITU after transposition of the specifications in a Standards Development Organization (SDO) into a standard and submittal via the national process, as applicable, into the ITU. For more information, see "*3rd Generation Partnership Project 2*". (accessed August 2001) Link: http://www.3gpp2.org/.

## 4.3   Is 3G Unique?

### 4.3.1   The Heavy Burden of the 3rd Generation – Consolidation Trends

Several aspects of IMT-2000 (UMTS) deployment and license allocation make the story of 3G development unique, particularly where 'consolidation' and the fate of 'new entrants' are concerned.  Governments have had to consider the real possibility that operators will join forces in order to absorb the shock of paying for spectrum.  Surely, no such pattern was prevalent in the early 1990's.  What implications does this have for the original goal of creating a competitive, 'healthy' mobile telecommunications sector?

Already the first signs of major operator collaboration are discernible, as bidders struggle to limit commercial rivalry: last July, Dutch operator KPN embarked upon a joint venture with Japan's DoCoMo, and Hong Kong-based Hutchison Whampoa.  The huge cost of 3G has led the majority of operators to begin discussions with their domestic competitors to share networks in order to reduce build-out expenses.  In most countries like Sweden, Italy, Spain and the UK, regulators are open to this proposition.  In Germany, however, where the cost of 3G is higher than in any other country, the regulator is still opposed to network sharing.[123]  The market is currently rife with such news of operators taking varying percentage equity stakes in international counterparts around the globe; this reflects the ceaseless activity of strategic positioning and re-positioning in the quest to attain scaled economies and balanced 'product' and 'service offering' 'portfolios'.

In the meantime, new entrants like Group 3G in Germany and TIW/Hutchison in the United Kingdom will not be able to alone withstand the pressures of having no organic customer base and vast cost disadvantages vis-à-vis incumbents.  These financial considerations indicate that by 2005, consolidation will have continued in Europe and operators will have coalesced or aligned themselves into fewer, larger groups. Efficient operators in one country will be able to improve the cost structures of smaller operators in other countries.  So far, Vodafone, as an outsider, has been among the boldest and most successful in terms of becoming a pan-European operator.  New entrants in this sector, according to some extremely pessimistic analysts, are doomed before they even begin; Forrester, for example, expects no new UMTS entrants to be left standing by the year 2007.[124]  However, given the reality of success stories like Vodafone and Mannesman, this assertion is questionable.  In any case, incumbent operators appear - not unlike in their old monopoly days - to be curiously well-placed to make some money (regardless whether or not a few new entrants survive).

### 4.3.2   3G Deployment Costs

3G operators, aside from worrying about the spectrum licenses, must invest in building or expanding their physical infrastructures.  Infrastructure is and continues to be a primary concern in the realm of estimated operator costs, and one common assumption is that these costs will come close to the amounts that operators paid for their spectrum licenses.  The truth is that a lot of people do not know how much building a UMTS network is going to cost, and estimates range widely.  The Yankee Group, for one, estimates average roll-out (including license, network infrastructure, application and content development) costs at $2.5 billion.[125]  An article in Red Herring from September 1, 2000 cites that building out infrastructure for 3G services is

---

[122] A good example of this type of collaboration between operators, manufacturers and content providers is evident in the case of Japan.  Equipment manufacturers and operators work hand in hand in closely-knit groups to supply the market with handsets and portable devices in line with end-user needs.  The mobile operator actually owns the handsets. As such, the operator's brand is dominant and not the manufacturer's. The Japanese subscriber first selects the service provider and then chooses the equipment, and the subscriber's choice of handset is therefore limited to those on offer and branded by the service provider selected.  This differs greatly from the European case, where the handset brand rests firmly with manufacturers such as Nokia and Ericsson, as does the responsibility for research and development.  Japanese mobile operators also play a leading role in research and development activities.  See Section 4.4 of the <u>International Telecommunication Union</u> 3G Japan case study, located at Link: http://www.itu.int/osg/spu/ ni/3G/casestudies/japan/_Toc523133746. (accessed September 2001)

[123] Both the EU and the German 3G license holders are exerting pressure on regulator, and Yankee Group expects that the regulator will bow to that pressure and concede that network sharing is a necessary step for the success of the German 3G market.

[124] Forrester also predicts that operating profits will disappear in 2007 and take six years to return, leading to major operator business failures and massive industry consolidation.  Godell, Lars. "*Europe's UMTS Meltdown*".  <u>Forrester Research Report</u>, December 2000, p.15.

[125] "*Wireless: riding its luck into 3G*".  <u>Mobile Matters</u>, February 2001, p. 52.

estimated in the ballpark of around $5 billion per operator per country.[126]   One operator interviewed by Forrester cites an internal estimate at around US$7.3 billion, although others think it may well cost US$4.5 billion beyond that.[127]   In the United Kingdom, Vodafone paid nearly US$8.6 billion for its license and then signed up Ericsson to provide the infrastructure in a deal understood to be worth around US$5.8 billion.   Similarly, in Germany, Mannesmann committed to a DM10 billion (US$4.6 billion) network upgrade after paying about DM16 billion (US$7.5 billion) for the 3G license.[128]   Looking at a typical such UK operator, it is estimated that cumulative costs of approximately $10 billion would be necessary in preparation for data services.[129]   In essence, estimates are numerous and not necessarily in the same ballpark.

IMT-2000 terminals for mass market sale will not be available in volumes until 2003 at the earliest, pushing back expected service launches at least a year or so.   Even in a country like Norway with relatively inexpensive licenses and networks builds, Telecom consultancy Teleplan states that ARPU must double by 2004 to recover UMTS costs.[130]   Many believe that long before 3G networks are completed, alternative solutions – such as the intermediate 2.5G technologies mentioned above - could replace them.

In any case, spending on IT systems and billing is likely to change significantly for the worse.  It is expected that network operators will have to take into account changes in their cost structures, particularly compared to their past experience with GSM systems.[131]   Amidst strong competition and high customer churn rates, Forrester predicts that marketing costs in particular will increase significantly before subsiding again after 2008.  Money spent on customer retention will be crucial in order to combat the churn 'whirlpool'.  Thus, high capital investment is unlikely to subside, as location-sensing equipment, content distribution facilities, and IP routers will incent operators to keep up with the pace of change.

Certainly, the cumulative costs of building a UMTS network will be different for incumbents and for new entrants.   "New entrants with no existing [GSM] infrastructure to reuse face … [high] costs, estimated at US$6.2 billion, as in the case for the Sonera/Telefonica alliance in Germany."[132]   Regulators, recognizing the ferocity of the marketplace particularly for non-incumbents, are requiring transition periods in which new entrants can use some of the incumbents' infrastructures while constructing their own.   Gartner Group analysis of the mix of operators' costs yields a hypothesis that the marginal costs of servicing a thousand new subscribers will rise from an average of $200 per subscriber for a GSM network to around $350 per subscriber on a UMTS network. (See Table 4.1)  Physical infrastructure costs will shrink from 65% of the total to 59% as other costs double.

**Table 4.1:  Estimated cost of GSM and UMTS networks**

| | Cost per Subscribers | | Percent Change | GSM (percent) | UMTS (percent) |
|---|---|---|---|---|---|
| | GSM | UMTS | | | |
| Core Network | $20.00 | $24.50 | 22.5% | 10 | 7 |
| Radio Network | $70.00 | $101.50 | 45.0% | 35 | 29 |
| Transmission Links | $40.00 | $80.50 | 101.3% | 20 | 23 |
| Network Maintenance | $22.00 | $38.50 | 75.0% | 11 | 11 |
| Sales and Marketing | $16.00 | $35.00 | 118.8% | 8 | 10 |
| Customer Care and Billing | $20.00 | $42.00 | 110.0% | 10 | 12 |
| IT Management Services | $12.00 | $28.00 | 133.3% | 6 | 8 |
| **Total** | **$200.00** | **$350.00** | **75.0%** | **100** | **100** |

Source: *Gartner Dataquest*

---

[126] Cukier, Kenneth and Hibbard, Justin.  "*Spectrum Shortage*".  Red Herring Magazine, September 1, 2000.

[127] Godell, Lars. "*Europe's UMTS Meltdown*".  Forrester Research Report, December 2000,  p.5.

[128]   Bout, Dirk M., Daum, Adam, Deighton, Nigel, Delcroix, Jean-Claude, Dulaney, Ken, Green-Armytage, Jonathan, Hooley, Margot, Jones, Nick, Leet, Phoebe, Owen, Gareth, Richardson, Peter, Tade, David.  "*The Next Generation of Mobile Networks Poses a $100 Billion Challenge for Europe*", Note Number: R-11-5053, Gartner Group.  September 19, 2000.

[129]  This would include US$6.3 billion on acquiring a 3G license, US$3 billion on building the 3G network, US$75 million on upgrading existing networks to GPRS and the remainder on content and service creation.  Bratton, William, Jameson, Justin, and Pentland, Stephen. "*Analysis: 3G madness – time for some moderation!*" Totaltele.com, July 16, 2001,  p.2.

[130] Godell,  p.8.

[131]   According to the Gartner Group, mobile services are moving from hierarchical architectures based on circuit switching, to distributed and layered architectures based on packet routing.  As a result, it is estimated that infrastructure costs may not increase in fact as much as other support costs in the long-run.  However, heavy investment in network management, billing systems, massive marketing, support services and handset subsidies are an inevitable part of the future.  Costs of billing systems in particular will rise sharply, since 'always-on' services will very likely disallow the relevance of per-minute charging.

On the other hand, extrapolation leads to an interesting scenario, wherein there is significant potential for a good turn of operators' fortunes after all. "Assuming an operator, which has 9 million subscribers between now and the end of 2020, starts to generate data revenues from only 2003, and experiences only a slow increase in ARPUs from US$2 per subscriber per month in 2003 to US$25 (in nominal terms), in 2012, then the present value of the expected revenues from data services, using a discount rate of 12%, is approximately US$11.5 billion. Not only does this cover the initial [aforementioned] US$10 billion of expenditure, but the assumptions are conservative."[133]

The question still remains, however, how operators are going to manage their interest payments on the $300 billion that will be sunk into European licenses and equipment. With a rate of 7%, operators still are going to need to earn $21 billion a year - just to pay interest. [134] Which brings us back, inevitably, to basic questions surrounding mobile penetration: how certain is it that increasingly complex data services - available via 2.5G and 3G networks and systems- will have what it takes to live up to global expectations?

# 5    Conclusion

To a large extent, GSM can be said to have been "the right system at the right place at the right time"[135]. Based on the analysis of this paper, it appears that the essence of the GSM story revolves around the concept of cooperation, and the political and economic environment that facilitated it. A main theme throughout this paper is that investments in the respective IMT-2000 standards are extremely high, and that those sustaining these commitments consist of a number of highly leveraged stakeholders like manufacturers, distributors, and standards consortia – all keen to justify their own paths toward IMT-2000. While European Community policy and Commission leadership were indispensable for GSM, flexibility and adaptability on the national level were vital for success. This is one of the key differentiating factors between the developments of 2[nd] generation and 3[rd] generation technologies.

In the broadest scope, the transition from 2G to 3G would have been inconceivable had it not been for the justifications of significant forecasted increases in mobile penetration numbers for the coming years. Certainly, the impact of tremendous network externalities is at its very core associated with this growth potential.

We have seen that IMT-2000 has been a rocky road because the multitude of players that will benefit from its deployment (including governments) have stood in fact to gain more individually (or even regionally) from compromising 'global standards harmonization' than from smoothly cooperating. The "Stag Hunt" example of game theory application reflects the dynamic of this scenario, in which players face a choice between finding a compromise and realizing the gains from collective action (i.e., closing in around a large target), or maintaining their positions as individual entities and accepting the risk (and with it, the potentially larger gains) associated with running alone after another target. Allowing two incompatible standards like W-CDMA and cdma2000 to come about was the result of the breakdown of the 'mutually beneficial collective action' mentality that characterized the decision-making dynamic of European nations as they constructed GSM.

Ironically, what is in evidence today in the 3G market, is the growth of consolidation and collaboration between operators that has made 3G unique from GSM. At earlier stages of IMT-2000 development, such collaboration (i.e., in the form of network-sharing) was unthinkable, given the widely divergent stakes in differing types of 2G legacy investments/commitments, and the general emphasis on the necessity for full and free competition in the marketplace. Perhaps this phenomenon of consolidation is reflective of a natural tendency for 'natural cooperation' intrinsic to the success of the mobile sector (i.e., the same way that monopolistic tendencies are 'natural' for the 'local-loop'), and that must inevitably emerge in some form as industry experience lead to it again and again. In this case, cooperation is perhaps being spurred by the workings of the spectrum market, despite the fact that the ability of political entities to bring it about was diminished.

---

[132] Godell, p.7.

[133] Bratton, William, Jameson, Justin, and Pentland, Stephen. "*Analysis: 3G madness – time for some moderation!*" Totaltele.com, July 16, 2001, p.2.

[134] Van Grinsven, Lucas. "*Mobile & Satellite: Nokia 3G guru cites SMS as key to wireless web success*". Reuters, June 28, 2001.

[135] Bach, p.1.

Although there has been some overlap in the time development of 2G and 3G, various key differences in the political and private sector catalysts for cooperation during the development of GSM have rendered the likelihood of replication of previous standardization achievements on a global scale not guaranteed. The value of the MoU, for example, as a beacon for industry governance in the GSM case, has not been imitable in global telecommunications fora. Incentives for global cooperation towards the creation of a uniform standard for 3$^{rd}$ generation technologies have been proven to be lacking, as is evident by the current wide range of IMT-2000 3$^{rd}$ generation 'flavours' (although the ITU has made significant achievements in the realm of standardization). Undoubtedly, the lack of consensus regarding harmonization across various IMT-2000 technologies is at least in part the result of 'cdma2000-oriented' policy objectives of concerned (mostly non-European) lobbyists. And although the European Commission has pursued and facilitated continuing collaboration between prominent players from the past decade, the fact remains that many conditions that helped render GSM a success simply no longer exist. Thus, while certain 'lessons' from the past can be applied to generate some value in terms of appraisal, again there is no guarantee that the actual events that characterized GSM's success will work for IMT-2000.

In an environment characterized by very rapid change and unmatched dynamism, it is an interesting task to pick and choose those factors that can be drawn in parallel from the past, to explain what is to come in the (albeit immediate) future. Though the political roots of GSM have transformed significantly into more market-driven ones for IMT-2000, we have seen that certain trends, metrics, and concepts are still relevant, while others fade into the background. One can but hope that past experience – as from the GSM case - breeds the types of institutions and leaders that are willing to learn from their mistakes and improve even further upon their successes as they seek to serve a global market.

# 6  Appendix:

**Table 6.1:  Allocation of 3G mobile licences in the European Union**

| Country | No of licences | Mobile Incumbents | Method | Date awarded | Sum paid US$ million |
|---------|----------------|-------------------|--------|--------------|----------------------|
| Austria | 6 | 4 | Auction | November 2000 | 610.0 |
| Australia | 6 | 4 | Auction | March 2001 | 351.7 |
| Belgium | 4 | 3 | Auction | February 2001 | 418.8 |
| Canada | 5 | 4 | Auction | January 2001 | 1,482.0 |
| Denmark | 5 | 4 | Auction | October 2001 | |
| Finland | 4 | 3 | Beauty contest + nominal fee | March 1999 | Nominal |
| France | 4 (2 still to be issued) | 3 (2 still to be issued) | Beauty contest + fee | July 2001 | 4,520.0 |
| Germany | 6 | 4 | Auction | July 2000 | 45,870.0 |
| Greece | 4 or more | 3 | Auction | July 2001 | |
| Ireland | 4 | 3 | Beauty contest + fee | April 2001 | Estimated between 116.0 and 140.0 each |
| Italy | 5 | 4 | Auction | October 2000 | 10,070.0 |
| Korea | 3 | 2 | Beauty Contest + fee | End 2000 | 3,080.0 |
| Luxembourg | 4 | 2 | Beauty Contest | By June 2001 | |
| Netherlands | 5 | 5 | Auction | July 2000 | 2,508.0 |
| New Zealand | 4 | 2 | Auction | January 2001 | 51.4 |
| Norway | 4 | 2 | Beauty contest + fee | November 2000 | 44.8 |
| Portugal | 4 | 3 | Beauty contest + fee | December 2000 | 360.0 |
| Spain | 4 | 3 | Beauty contest + fee | March 2000 | 120.0 each |
| Sweden | 4 | 3 | Beauty contest | December 2000 | 44.08 |
| Switzerland | 4 | 2 | Auction | December 2000 | 116.0 |
| UK | 5 | 4 | Auction | April 2000 | 35,390.0 |

Source: ITU, European Commission, The Introduction of 3G Mobile Communications in the European Union: State of Play and the Way Forward, Brussels 20.3.2001 COM(2001)141final..